\documentclass[fleqn,usenatbib]{mnras}
\usepackage{newtxtext,newtxmath}
\usepackage[T1]{fontenc}
\usepackage{ae,aecompl}
\usepackage{graphicx}	
\usepackage{amsmath}	
\usepackage{amssymb}
\usepackage{hyperref}

\DeclareMathOperator{\sech}{sech}

\title[A multi-messenger study of the Milky Way]{A multi-messenger study of the Milky Way's stellar disc and bulge with LISA, {\sl Gaia} and LSST}

\author[V. Korol et al.]{
Valeriya Korol,$^{1}$\thanks{E-mail: korol@strw.leidenuniv.nl}
Elena M. Rossi,$^{1}$
Enrico Barausse$^{2}$
\\
$^{1}$Leiden Observatory, Leiden University, PO Box 9513, 2300 RA, Leiden, the Netherlands\\
$^{2}$Institut d'Astrophysique de Paris, CNRS \& Sorbonne Universit{\'e}s, UMR 7095, 98 bis Bd Arago, 75014 Paris, France\\
}

\date{Accepted XXX. Received YYY; in original form ZZZ}
\pubyear{2018}

\begin{document}
\label{firstpage}
\pagerange{\pageref{firstpage}--\pageref{lastpage}}
\maketitle

%%%%%%%%%%%%%%%%%%%%%%%%%%%%%%%%%%%%%%%%%%%%%%%%%%%%%%%%%%%%%%%%%%%%%%%%
\begin{abstract}
The upcoming LISA mission offers the unique opportunity to study the Milky Way through gravitational wave radiation from a large population of Galactic binaries.
Among the variety of Galactic gravitational wave sources, LISA is expected to individually resolve signals from  $\sim 10^5$ ultra-compact double white dwarf (DWD) binaries. DWDs detected by LISA will be distributed across the Galaxy, including regions that are hardly accessible to electromagnetic observations such as the inner part of the Galactic disc, the bulge and beyond. 
We quantitatively show that the large number of DWD detections will allow us to use these systems as tracers of the Milky Way potential. 
We demonstrate that density profiles of DWDs detected by LISA may provide constraints on the scale length parameters of the baryonic components that are both accurate and precise, with statistical errors of a few percent to $10$ percent level. Furthermore, the LISA sample is found to be sufficient to disentangle between different (commonly used) disc profiles, by well covering the disc out to sufficiently large radii.  
Finally, up to $\sim 80$ DWDs can be detected  through both electromagnetic and gravitational wave radiation. 
This enables multi-messenger astronomy with DWD binaries and allows one to extract their physical properties using both probes.
We show that fitting the Galactic rotation curve constructed using distances inferred from gravitational waves {\it and} proper motions from optical observations yield a unique and competitive estimate of the bulge mass. Instead robust results for the stellar disc mass are contingent upon knowledge of the Dark Matter content.
\end{abstract}
%%%%%%%%%%%%%%%%%%%%%%%%%%%%%%%%%%%%%%%%%%%%%%%%%%%%%%%%%%%%%%%%%%%%%%%%%%

\begin{keywords}
gravitational waves -- white dwarfs -- binaries:close -- Galaxy:structure -- Galaxy:bulge -- Galaxy:disc
\end{keywords}

%%%%%%%%%%%%%%%%%%%%%%%%%%%%%%%%%%%%%%%%%%%%%%%%%%

%%%%%%%%%%%%%%%%% BODY OF PAPER %%%%%%%%%%%%%%%%%%

\section{Introduction}

Because of our vantage observation point, the Milky Way is an outstanding laboratory for understanding galaxies, whose assembly histories bear the imprint of the cosmological evolution of our Universe.
As remnants of the oldest stars in the Milky Way, white dwarfs (WDs) are unique tracers of the Milky Way's properties.
For example, using the fact that the WD luminosity depends mainly on the stellar age, one can date different Galactic populations by constructing a WD luminosity function \citep{Liebert1988,Rowell2011,GB2016,Kilic2017}.
Moreover, the WD luminosity function contains information about the star formation and death rates over the history of the Galaxy.
The most ancient WDs in the Galaxy can make up a sizeable fraction of the dark Galactic stellar halo mass, and, thus, have a direct impact on our quantitative estimates of the total amount of dark matter in the Galaxy \citep[e.g.][]{Alcock2000, Flynn2003, Napiwotzki2009}. 
In this work we quantitatively show that WDs in close binaries are unique multi-messenger tools to probe the Milky Way's structure.

Double WDs (DWDs) are expected to be detected through gravitational wave (GW) emission by the Laser Interferometer Space Antenna (LISA), an ESA space mission officially approved in 2017 \citep{LISA2017}.
LISA is designed to detect GW sources in the mHz frequency range, such as merging massive black hole binaries ($\sim 10^{4}\,$M$_{\odot}-10^{7}\,$M$_{\odot}$) up to $z\sim15-20$ \citep[e.g.][]{Klein2016}, extreme mass ratio inspirals \citep[e.g.][]{Babak2017} and Galactic binaries \citep{Korol2017,Kremer2017,Breivik2018}. Therefore, besides  probing high-redshift cosmology \citep{Caprini2016,Tamanini2016} and testing the theory of General Relativity in the strong gravity regime \citep{Barausse2016,Berti2016,Brito2017}, LISA will be the only gravitational experiment capable of exploring the Milky Way's structure.
Remarkably, the expected number of Galactic binaries that LISA will be able to resolve individually (i.e. measure their individual properties) amounts to $\sim 10^5$, among which DWDs will represent the absolute majority \citep[e.g.][]{Nelemans2004,Ruiter2010,Shah2012,Kremer2017,Korol2017}. Overlapping signals from unresolved binaries present in the Galaxy will instead form a stochastic background signal \citep{Edlund2005,Timpano2006,Robson2017}. 
Both resolved and unresolved LISA signals will provide information on the Galactic stellar population as a whole, and can thus be used to study the Milky Way's baryonic content and shape.
A first quantitative study was carried out by \citet{Ben2006}, where the authors show that the level and shape of the DWD background as well as the distribution of resolved sources will provide constraints on the scale height of the Galactic disc. In this paper we focus on resolved binaries only and we demonstrate their potential for constraining the shape of both the disc and the bulge. Moreover, we show that the power to constrain the overall properties of the Galactic baryonic potential will be significantly enhanced by using GWs in combination with electromagnetic (EM) observations. The success of this synergy is due to LISA's ability to localise binaries through virtually the whole Galactic plane, thus mapping its shape, while optical observations yield the motion of stars, tracing the underlying total enclosed mass.

In this work, we use a synthetic population of detached DWD binaries (Section~\ref{sec:population}) to investigate the precision of LISA distance measurements (Section~\ref{sec:distances}) and to test the potential of using the spatial distribution of the LISA detections to reconstruct the density profiles of the Milky Way stellar population (Section \ref{sec:4}). We focus on detached binaries because they are ``clean'' systems where systematics in the system's parameter determination are reduced. We also simulate the performances of {\sl Gaia} and the LSST at providing astrometric measurements for eclipsing binaries, and we simultaneously fit the stellar density shape {\it and} the Milky Way's rotation curve (Section \ref{sec:5}). In Section \ref{sec:6} we present our conclusions.

\section{Synthetic population}
\label{sec:population}
The detailed description of our population synthesis model was presented in \citet{Toonen2012,Toonen2017} and \citet{Korol2017},  to which we refer for further details.
In this section we summarise the most important features of the adopted model, focussing on the Milky Way structure and potential.
We also outline the method that we have used to simulate detections of DWDs with {\sl Gaia} and the LSST, and the computation of the signal-to-noise ratios for the latest design of the LISA mission \citep{LISA2017}.

\subsection{Initial distributions}

In modelling the synthetic population of DWDs we rely on the population synthesis code SeBa \citet[][for updates see \citealt{Nelemans2001a}, \citealt{Toonen2012}]{SeBa}.
The initial stellar population is obtained with a Monte Carlo approach, assuming a binary fraction of 50\% and adopting the following distributions for the binary parameters.
First, we draw the mass of the single stars between 0.95 - 10 M$\odot$ from the Kroupa initial mass function \citep[IMF,][]{KroupaIMF}.
Then, we draw the mass of the secondary star from a flat mass ratio distribution between 0 and 1 \citep{Duchene2013}.
We adopt  a log-flat distribution for the binary semi-major axis and  a termal distribution for the orbit eccentricity \citep{Abt1983,Heggie1975, Raghavan2010}.
Finally, we draw the binary inclination angle $i$ isotropically (i.e. from a uniform distribution in $\cos i$).
The sensitivity of our population model to these assumptions is discussed in \citet{Korol2017} and \citet{Toonen2017}.

In the canonical picture of binary evolution, a common envelope (CE) phase is required   to form a close system \citep{Paczynski1976,Webbink1984}.
This is a short phase in binary evolution in which the more massive star of the pair expands and engulfs the companion.
When this happens the binary orbital energy and angular momentum can be transferred to the envelope, due to the dynamical friction that the companion star experiences when moving through the envelope.
Typically, this process is implemented in the binary population synthesis by parametrising the conservation equation for either the energy or the angular momentum \citep[see][for a review]{Ivanova2013}.
In our previous work we modelled two populations, one for each CE parametrisation, to study whether optical surveys such as  {\sl Gaia} or LSST, as well as LISA in GWs, will be able to discriminate between the two.
In this paper we are mainly interested in the spatial distribution of DWDs, which does not depend on the specific CE prescription, and thus we use just one model population.
In particular, we choose the parametrisation based on the angular momentum balance ($\gamma$-parametrisation), which was introduced to reconstruct the population of observed DWDs and was fine-tuned using them \citep{Nelemans2000, Nelemans2001a, Nelemans2005}.

\subsection{Galaxy model: density distribution, potential and rotation curve}

%%%%%%%%%%%%%%%%%%%%%%%%%%%%%%
\begin{table} \label{tab:1}
\begin{center}
\caption{Milky Way model}
\label{tab:1}
\centering
 \begin{tabular}{ l | c | r }
 \hline
   Parameter & Value \\ \hline \hline
   \multicolumn{2}{c}{Bulge} \\
   $M_{\rm b}$ & $2.6 \times 10^{10}\,$M$_{\odot}$  \\ 
   $r_{\rm b}$ & $0.5\,$kpc  \\
   $r_{\rm b,max}$ & $3\,$kpc  \\ \hline
   \multicolumn{2}{c}{Stellar disc} \\
   $M_{\rm d}$ & $5 \times 10^{10}\,$M$_{\odot}$  \\ 
   $R_{\rm d}$ & $2.5\,$kpc  \\ 
   $R_{\rm d,max}$ & $19\,$kpc  \\ 
   $Z_{\rm d}$ & $0.3\,$kpc  \\  \hline
   \multicolumn{2}{c}{DM halo} \\
   $\rho_h$ &  $0.5 \times 10^7\,$M$_{\odot}$kpc$^{-3}$ \\
   $M_{\rm h}$ & $4.8 \times 10^{11}\,$M$_{\odot}$  \\ 
   $r_{\rm h}$ & $20\,$kpc  \\ 
   $r_{\rm h,max}$ & $100\,$kpc  \\ \hline
  \end{tabular}
\end{center}
\end{table}
%%%%%%%%%%%%%%%%%%%%%%%%%%%%%

%%%%%%%%%%%%%%%%%%%%%%%%%%%%%
\begin{figure}
        \centering
	 \includegraphics[width=0.5\textwidth]{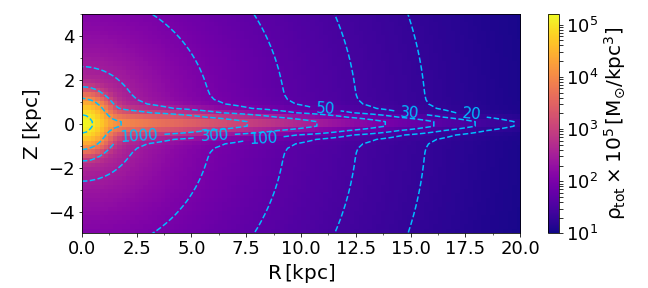} 
	 \includegraphics[width=0.5\textwidth]{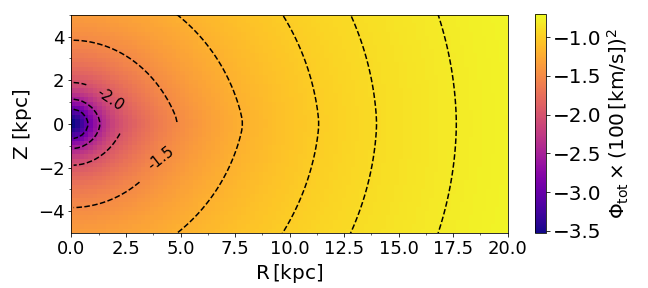} 
         \caption{Density and potential maps of our Milky Way fiducial model in the $R-Z$ plane, computed numerically with the {\sc galpynamics} package. Contour levels in the upper panel are $(20, 30,50,100, 300, 10^3, 10^4, 10^5) \times 10^{5}\,$M$_{\odot}$/kpc$^3$. Contour levels in the lower panel corresponds to $(-3, -2.5, -2, -1.5, -1.2, -1, -0.9, -0.8)\times (100\,$km/s$)^2$.}
       \label{fig:MWpotential}
\end{figure}

\begin{figure}
        \centering
	 \includegraphics[width=0.45\textwidth]{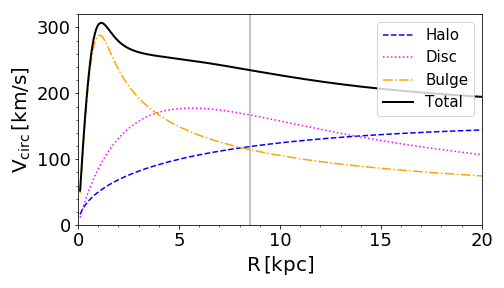} 
         \caption{Rotation curve of our Milky Way fiducial model. The contributions from the disc, bulge and halo are shown by the dotted magenta, dashed-dotted yellow and dashed blue curves respectively. The total circular speed, given by the sum in quadrature of the circular speeds of the components, is represented by the black solid line. The circular velocity at the position of the Sun ($8.5\,$kpc), marked by the grey vertical line, is $235\,$km/s.} \label{fig:MWrotcurve}
\end{figure}
%%%%%%%%%%%%%%%%%%%%%%%%%%%%%

We consider a simple model for the Milky Way, which we assume to be comprised of a bulge, a stellar disc and a dark matter (DM) halo.
We distribute DWDs in the bulge and in the disc, while the DM halo is needed to reproduce Galactic kinematics.
We do not take into account the stellar halo component because the properties of the WD population in the halo, and  those of the  stellar halo itself, are not well known \citep[e.g.][]{Cojocaru2015}.
Furtermore, the signal arising from the halo population is not expected to contribute significantly to the overall GW signal from the Galaxy \citep{Ruiter2009}.

The density of DWDs in the disc is assumed to fall exponentially in the radial direction, $R$, and to depend on the distance from the mid-plane, $Z$, through a $\sech^2$ function \citep[e.g.,][]{Robin2014}.
For simplicity, we neglect the dependence on the stellar age and mass when distributing DWDs in the $Z$ direction, and we assume that they do not migrate radially.
To account for the star formation history of the Milky Way disc we use the plane-projected star formation rate from \citet{BoissierPrantzos1999}, $\rho_{\rm BP}$, and  assume the age of the Galaxy to be 13.5 Gyr \citep[e.g.][]{Juric2008}.
Analytically, the density distribution of the disc component for our model can be written as 
\begin{equation}  \label{eqn:disc}
\rho_{\rm disc}(t,R,z) = \rho_{\rm BP}(t)\, e^{-R/R_{\rm d}}\sech^2 \left( \frac{z}{Z_{\rm d}} \right) \ {\text M}_{\odot}\, {\text{kpc}}^{-3},
\end{equation} 
where $0 \le R \le 19$ kpc is the cylindrical radius measured from the Galactic centre, $R_{\rm d} = 2.5\,$kpc is the characteristic scale radius, and  $Z_{\rm d}=300$ pc is the characteristic scale height of the disc \citep{Juric2008}. 
The total mass of the disc in our model is $5 \times 10^{10}\,$M$\odot$.
We assume the distance of the Sun from the Galactic centre to be $R_{\odot}=8.5\,$kpc \citep[e.g.][]{2012Schondrich}.

We model the bulge component by doubling the star formation rate in the inner $3\,$kpc of the Galaxy and distributing DWDs according to 
\begin{equation} \label{eqn:bulge}
\rho_{\rm bulge}(r) = \frac{M_{\rm b}}{(\sqrt{2\pi} r_{\rm b})^3} e^{-r^2/2 r_{\rm b}^2}\  {\text M}_{\odot}\, {\text{kpc}}^{-3},
\end{equation}
where $r$ is the spherical distance from the Galactic centre, $M_{\rm b} =  2.6 \times 10^{10}\,$M$\odot$ is the total mass at the present time, and $r_{\rm b}=0.5\,$kpc is the characteristic radius \citep[e.g.][]{Sofue2009}.

To model the density distribution of the DM halo we use the Nawarro-Frenk-White profile \citep{NFW1996}:
\begin{equation} \label{eqn:halo}
\rho_{\rm DM} (r) = \frac{\rho_{\rm h}}{(r/r_{\rm s})(1+r/r_{\rm s})^2} \ {\text M}_{\odot}\, {\text{kpc}}^{-3},
\end{equation}
where $r_{\rm s} = 20\,$kpc is the scale length of the halo and $\rho_{\rm h} = 0.5 \times 10^7\,$M$_{\odot}$kpc$^{-3}$ is the halo scale density.
The total mass of the halo can be obtained by integrating eq.~\eqref{eqn:halo} from the centre to the maximum Galactocentric radius  of 100$\,$kpc, which for our fiducial parameters yields $4.8 \times 10^{11}\,$M$_{\odot}$.
We summarise the values of the  parameters adopted for our Milky Way fiducial model in Table \ref{tab:1}.

The total potential can be computed by solving the Poisson equation
\begin{equation}\label{eqn:poisson}
{\bf \nabla}^2 \Phi_{\rm tot} = 4 \pi G (\rho_{\rm disc} + \rho_{\rm bulge} + \rho_{\rm DM}).
\end{equation}
We solve eq.~\eqref{eqn:poisson} numerically using the {\sc galpynamics} Python package, which is designed for the computation and fitting of potentials, density distributions and rotation curves\footnote{{\sc galpynamics} is a free source Python package developed by G. Iorio and available at \url{https://github.com/iogiul/galpynamics}}.
We represent the resulting total density distribution and potential in Fig. \ref{fig:MWpotential}.
Both panels show a very prominent and concentrated bulge component reflected by the much closer iso-density (upper panel) and equipotential (lower panel) contour lines near the centre. 
The contribution of the disc inside the solar Galactocentric radius is clearly seen in the upper panel, and can be inferred from the flattening of the equipotential lines in the vertical direction in the lower panel of Fig. \ref{fig:MWpotential}.
At $R > 15\,$kpc the halo component becomes dominant, as reflected by the spherical shape of the iso-density and equipotential contours. 
We compute the Galactic rotation curve numerically using {\sc galpynamics} as
\begin{equation} \label{eqn:rot_curve}
V_{\rm circ}^2(R) = R\frac{{\rm d}\Phi_{\rm tot}}{{\rm d}R}.
\end{equation}
The result is illustrated in Fig. \ref{fig:MWrotcurve}, which
 shows that in our Milky Way model the bulge component has an important dynamical effect in the central region of the Galaxy up to $\sim 4\,$kpc.
In the region between 4 and 14$\,$kpc, the disc dominates the dynamics of the Galaxy, while at larger radii the DM halo provides the largest contribution to the rotation curve.
In our model the circular velocity at the position of the Sun is $V_0 = 235\,$km/s.
To compute the random component of DWD motion, 
we assume that the velocity distribution in the disc is governed by only two constants of motion, the energy and the angular momentum along the $Z$ direction.
Consequently, the specific low-order moments of the velocity components can be found as \citep[][]{binney2011galactic}
\begin{equation} \label{eqn:BT}
\overline {v^2_{\rm R}} = \overline{ v^2_{\rm Z}} = \frac{1}{\rho(R,Z)} \int_Z^{\infty} dZ' \rho(R,Z') \frac{\partial \Phi_{\rm tot}}{\partial Z'},
\end{equation}
where $\rho(R,Z)$ is the density  distribution of the Galactic component  (bulge or disc) in cylindrical coordinates.
Assuming that there is no stellar motion  in the radial and vertical directions, eq.~\eqref{eqn:BT} provides a direct estimate of the velocity dispersion $\sigma_{\rm R}$ and $\sigma_{\rm Z}$.
From eq.~\eqref{eqn:BT} we obtain the velocity moment in the azimuthal direction:
\begin{equation} \label{eqn:sigma_vphi}
\overline{v_{\phi}^2} = \overline{v^2_{\rm R}} + \frac{R}{\rho} \frac{\partial(\rho {\overline{v_{\rm R}^2}})}{\partial R} + R\frac{\partial \Phi_{\rm tot}}{\partial R}.
\end{equation}
We evaluate the last two equations numerically using {\sc galpynamics}.
At the Sun's position we obtain $\sigma_{\rm R}, \sigma_{\phi}$ and $\sigma_{\rm Z}$ equal to $15,30$ and $15\,$km/s respectively.

\subsection{WD magnitudes}

The absolute magnitudes of WDs (bolometric and {\it ugriz}-Sloan bands) in our simulation are calculated from the WD cooling curves of pure hydrogen atmosphere models \citep[][and references therein]{Holberg2006,Kowalski2006,Tremblay2011}.
To convert the absolute magnitudes to observed magnitudes (e.g. for the Sloan $r$ band) we use:
\begin{equation}
r_{\rm obs} = r_{\rm abs} + 10 + 5\log d + 0.84 A_{\rm V},
\end{equation}
where $d$ is the distance to the source in kpc, $0.84 A_{\rm V}$ is the extinction in the Sloan $r$ band and $A_{\rm V}$ is the extinction in the $V$ band.
To compute the value of $A_{\rm V}$ at the source position, defined by the Galactic coordinates $(l,b)$ and the distance $d$, we use
\begin{equation}
A_{\rm V} (l, b, d) = A_{\rm V} (l,b) \tanh \left(\frac{d \sin b}{h_{\rm max}} \right), 
\end{equation}
where $A_{\rm V} (l,b)$ is the integrated extinction in the direction defined by $(l, b)$ 
from \citet{Schlegel1998}, $h_{\rm max} \equiv \min( h, 23.5 \times \sin b)$ and $h = 120$ pc is the Galactic scale
height of the dust \citep{Jonker2011}.
To convert {\it ugriz}-magnitudes into {\sl Gaia} $G$ magnitudes we apply a colour-colour polynomial transformation with coefficients chosen according to \citet{Carrasco2014}.

%%%%%%%%%%%%%%%%%%%%%%%%%%%%%%%%%%%

\subsection{Detection of DWDs with LISA}

%%%%%%%%%%%%%%%%%%%%%%%%%%%%%%%%%%%
\begin{figure*}
        \centering
	 \includegraphics[width=0.9\textwidth]{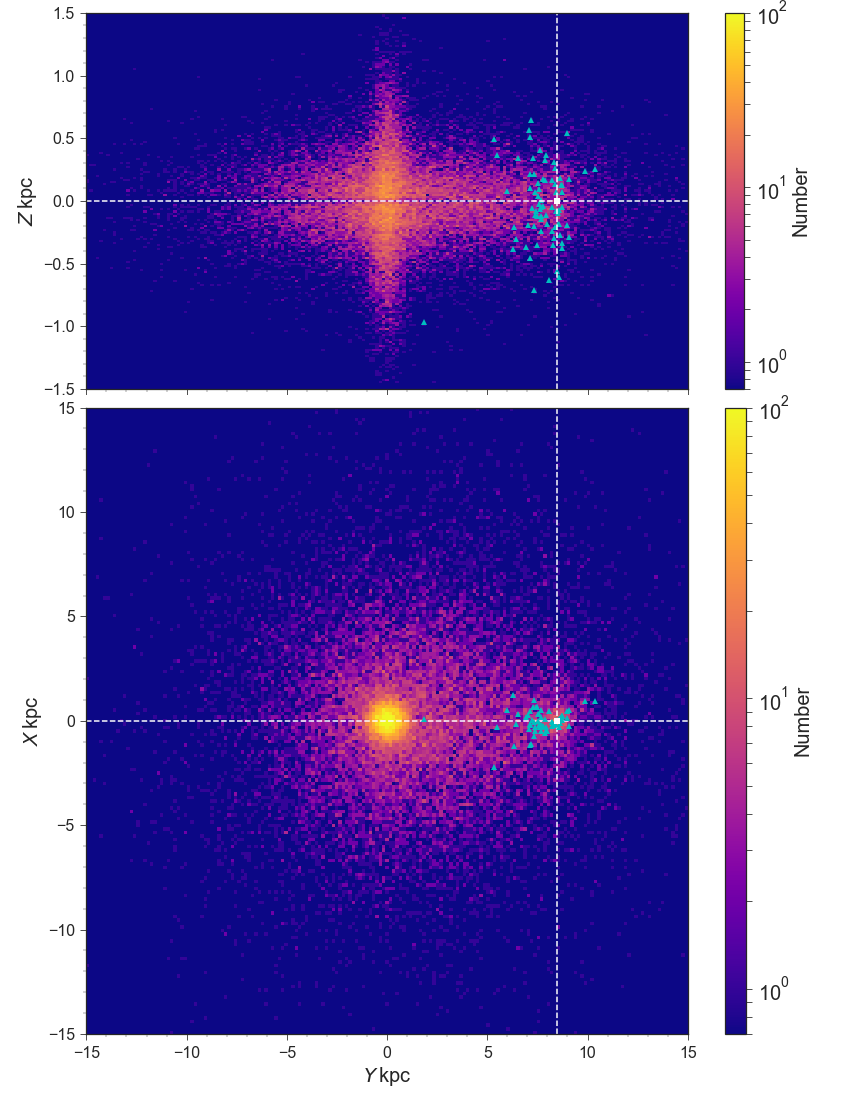}
         \caption{Source-count maps of DWDs detected by LISA (SNR>7) in the Galactocentric Cartesian coordinate system defined by eq.~\eqref{eqn:coord}: in the $Y-Z$ plane (top panel) and in the $Y-X$ plane (bottom panel). The white square identifies the position of the Sun in the Galaxy, $(0,8.5 {\rm kpc},0)$. Blue triangles represent the position of EM counterparts detected with {\sl Gaia} and/or LSST.}
       \label{fig:3} \label{fig:MWinGWs}
\end{figure*}
%%%%%%%%%%%%%%%%%%%%%%%%%%%%%%%%%%%%

GWs produced by a binary of  compact objects sufficiently far from coalescence at the lowest order can be described by the quadrupole approximation.
For a circular binary the quadrupole approximation yields a coalescence time due to GW emission of \citep{Maggiore}:
\begin{equation} \label{eqn:tau}
\tau \simeq 1 \,{\rm Myr} \  \left( \frac{P}{12\,{\rm min}} \right)^{8/3} \left( \frac{{\cal M}}{0.3\,{\rm M}_{\odot}} \right)^{-5/3}, 
\end{equation} 
where we use typical values for the binary orbital period $P$ and the chirp mass ${\cal M} = (M_1 M_2)^{3/5}/(M_1 + M_2)^{1/5}$ for our population \citep[][fig.~(13)]{Korol2017}.
Thus, a typical merger time for a DWD in our mock catalogue is of the order of Myr.
This is six orders of magnitude larger than the LISA mission lifetime, thus DWDs can be treated as quasi-monochromatic GW sources. 
The dimensionless GW amplitude can be found as
\begin{equation} \label{eqn:GWamp}
A = \frac{5}{96\pi^2}\frac{\dot{f}}{f^3d}
\end{equation}
where $f=2/P$ is the GW frequency,
\begin{equation} \label{eqn:GWfdot}
	\dot{f} = \frac{96}{5} \pi^{8/3} \left( \frac{G{\cal M}}{c^3} \right)^{5/3} f^{11/3}
\end{equation}
is the frequency derivative or chirp \citep{Maggiore}.
From eq.~\eqref{eqn:GWamp}-\eqref{eqn:GWfdot}, it follows that the distance can be determined directly by measuring the three GW observables $f, {\dot f}$ and $A$.
However, this is possible only for detached binaries whose dynamics is driven only by emission of GWs.
In the case of accreting DWDs (so-called AM CVns), the chirp contains components of astrophysical origin such as mass transfer or tides.
Consequently, the distance to these sources needs to be determined differently and requires additional EM observations \citep[e.g.][]{Breivik2018}.
Since this work deals with the possibility of mapping the Milky Way potential by GW observations, we focus on detached DWDs only.
A distinction between the two types of systems in the LISA data is possible based on the sign of $\dot{f}$: detached (AM CVns) systems are expected to have positive (negative) $\dot{f}$. This is due to the fact that the frequency of an AM CVn system decreases with time because of mass transfer, while the frequency of a detached system increases because of GW emission \citep[e.g.][]{Nelemans2004}.

When considering a space mission such as LISA, which is constantly in motion with changing speed and position with respect to a source in the sky, it is more convenient to work in the heliocentric ecliptic reference frame.
In this frame the coordinates of the source are fixed and the modulation of the GW signal in time is encoded in the detector response function \cite[e.g.][]{Cutler1998}.
We use the {\sc pyGaia}\footnote{In this paper we extensively use tools provided by {\sc pyGaia}, such as transformations between astrometric observables and transformations between sky coordinate systems, not only for simulating {\sl Gaia} data, but also as a general astronomical tool.}, a Python tool kit to transform the coordinates of DWDs from the galactic heliocentric frame to the ecliptic heliocentric frame (so that $r_{\rm ecl}=d$), and we define the LISA reference frame as 
\begin{equation}
\begin{aligned}
r &=   r_{\rm ecl} \\
\theta &= {\pi}/{2} - \arccos({z_{\rm ecl}}/{r_{\rm ecl}})  \\
\phi &= \arctan({y_{\rm ecl}}/{x_{\rm ecl}}).
\end{aligned}
\end{equation}
To compute signal-to-noise ratios (SNRs) for our mock population of DWDs over the nominal $4\,$yr mission lifetime, we employ the Mock LISA Data Challenge (MLDC) pipeline, which was designed for the simulation and analysis of GW signals from Galactic binaries \citep[for details see][]{Littenberg2011}.
The MLDC pipeline characterises  GW signals in terms of 9 parameters: $A, f, \dot{f}, \ddot{f}$, sky location ($\theta$, and $\phi$), orbital inclination $\iota$,  GW polarisation $\psi$ and the binary initial orbital phase $\phi_0$.
Given a synthetic instrument noise curve, and setting an observation time and a detection threshold, the MLDC pipeline provides a catalogue of the sources that can be resolved individually (i.e. those with SNR above the detection threshold), computes the background from unresolved sources in the catalogue, and estimates the uncertainties on the source parameters by computing the Fisher Information Matrix (FIM). We adopt the detector's design as approved by ESA, i.e. a three-arm configuration with $2.5\times 10^6\,$km arm length and the instrumental noise curve from \citet{LISA2017}.

We find $2.6 \times 10^4$ DWDs in our catalogue with SNR>7.
Their distribution in the Milky Way is represented in Fig. \ref{fig:MWinGWs}: the source-count map is shown in the $Y-Z$ plane (top panel) and in the $Y-X$ plane (bottom panel).
We denote the position of the Sun by a white square.
Figure \ref{fig:MWinGWs} reveals that LISA will detect DWD binaries to large distances, mapping also the opposite side of the Milky Way.
Both maps show a prominent peak in the central part of the Galaxy, due to the bulge, whereas the number of detected sources declines when moving outwards (up to $>15\,$kpc) from the centre, tracing the underlying disc stellar population.
The $Y-X$ map shows an asymmetry with respect to the $Y=0$  line due to an observation bias.
Indeed, because the amplitude and SNR of GW signals scale as $1/d$, nearby sources have stronger signals, and consequently there are more detected DWDs around the Sun.
We derive a correction factor to compensate for this bias in Appendix \ref{app:bias}.

\subsection{Detection of optical counterparts with {\sl Gaia} and LSST} \label{sec:EMdetections}
Additional information (such as the motion of DWDs) needed to constrain the Milky Way potential  cannot be extracted from GW data, but can be recovered from EM observations.  The sky localization of a source is typically poorly constrained by GWs, compared to optical observations. A typical position error for LISA is $\sim 10\,$deg, while a typical position error  for {\sl Gaia} is of the order of $\mu$as \citep{Gaia}. This makes it difficult to identify counterparts to GW sources in EM databases. In practice, in order to assemble a sample of optical counterparts, one possibility is to search in optical catalogues for  periodically variable sources  with a frequency and within an area on the sky matching those provided by LISA. To assess whether this is possible we focus on edge-on binaries, which allow for better parameter estimation with GWs and are easy to identify in optical as eclipsing. In particular, we consider two optical surveys, which by the time LISA is launched will be operational and which are expected to provide large stellar catalogues: {\sl Gaia} and the Large Synoptic Survey Telescope \citep[LSST,][]{LSST}. Our previous study shows that the deep magnitude limit of 21 for {\sl Gaia} and 24 for the LSST enables the detection of a significant fraction of the overall DWD Galactic population \citep{Korol2017}. Here below we summarise our method and results.

We simulate the optical light curves of DWDs detectable with LISA by computing the flux of a binary for a given orbital phase.
We consider spherically symmetric stars with uniform surface brightness, neglecting the limb darkening effect.
In this purely geometric model, we ignore the gravitational distortion of the stars and their mutual heating, which is justified given the small size of WDs and the roughly equal size of the binary components.
To evaluate the relative photometric error of a  single observation with {\sl Gaia} in the {\sl Gaia} $G$-band  we use:
\begin{equation}
\sigma_{\rm G} = 1.2 \times 10^{-3} (0.04895\tilde{z}^2+1.8633\tilde{z}+0.00001985)^{1/2},
\end{equation}
where $\tilde{z}=\max [10^{0.4(12-15)} , 10^{0.4(G-15)}]$ \citep{Gaia}.
To evaluate the expected photometric error of a single observation (as an example we use {\sl Sloan} r-band) with the LSST we use 
\begin{equation}
\sigma_{\rm r} = (\sigma_{\rm sys}^2 + \sigma_{\rm rand}^2)^{1/2},
\end{equation}
where $\sigma_{\rm sys}=0.005$ is the systematic photometric error, $\sigma_{\rm rand}^2 = (0.04-{\tilde \gamma})x+{\tilde \gamma} x^2$, $x=10^{(m-m_5)}$ is the random photometric error, and $m_5$ and ${\tilde \gamma}$ are respectively the $5\sigma$ limiting magnitude for a given filter and the sky brightness in a given band \citep{LSST}.
Finally, we apply a Gaussian noise to our synthetic light curves.

Next, we sample the light curves using the predicted {\sl Gaia} observations obtained with the {\it Gaia Observation Forecast Tool}\footnote{http://gaia.esac.esa.int/gost/}, which provides a list of times (in TCB, Barycentric Coordinate Time) for a given target in the sky.
We assign the initial orbital phase and sample the synthetic light curves with {\sl Gaia} observations, which we compute for each source individually.
To simulate the LSST sampling we use the anticipated regular cadence of 3 days over the nominal ten-year life span of the mission. 
In order to establish the detectability of the light curves,  we first verify whether the time sequence of simulated observations presents variability, by evaluating the $\chi^2$ for the observation sequence with respect to the average magnitude; and, second, we require a minimum number of observations to sample the eclipse phase ($\sim 3$\% of the total number of observations).
For each binary we compute 100 realisations of the light curve sampling by randomising over the initial orbital phase, and we define the probability of detection as the number of times the light curve was classified as detected out of 100.

We find 25 and 75 EM counterparts of the LISA sources with  respectively {\sl Gaia} and LSST, in agreement with our previous work \citep[][where, however, we simulate GW signals differently]{Korol2017}.
Since there is an overlap of 23 binaries between {\sl Gaia} and LSST detections, the total number of unique EM counterparts actually  amounts to 78.
We represent these sources with blue triangles in Fig. \ref{fig:MWinGWs}.
It is evident that there is a lack of EM detections in the disc plane and in the central bulge (i.e. at low Galactic latitudes) due to extinction effects.
The majority of EM counterparts will be detected at short distances compared to the extension of the stellar disc: within $2\,$kpc with {\sl Gaia} and within $10\,$kpc with the LSST.
Thus, we anticipate that combined GW and EM catalogues will provide information mainly on the local properties of the Milky Way.
%%%%%%%%%%%%%%%%%%%%%%%%%%%%%%%%%%%%%%%%%%%%%%%%%

\section{Distance determinations}
\label{sec:distances}

The precise determination of distances is a crucial step for studying the spatial distribution of DWDs in the Galaxy.
For DWD binaries the distance can in principle be independently measured from GW and optical observations, when both are available.
In this section we first forecast the LISA performance at measuring distances when considering a 4-year long observation run, and then we turn to the distance determination from parallax with {\sl Gaia} and the LSST end-of-mission performances.
Finally, for the DWDs with optical counterparts, we show that parallaxes can be used to improve the GW distance estimates.
In the following we denote the distance estimated from GWs and its error with the subscript ``GW'', and  the distance estimated from parallax measurements and its error with the subscript ``EM''. As in previous Sections, we refer to $d$ with no subscript  as the true distance to the source.

\subsection{Distances from GW data}

%%%%%%%%%%%%%%%%%%%%%%%%%%%%%
\begin{figure} 
        \centering
	 \includegraphics[width=0.46\textwidth]{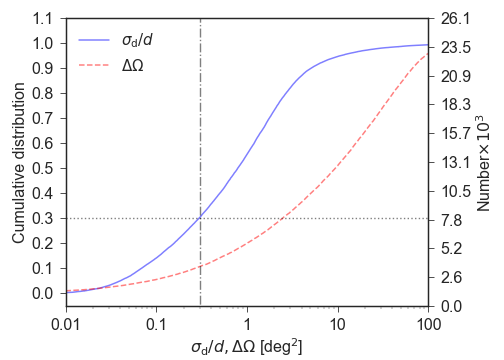} 
         \caption{Cumulative distribution (left y axis) and total number of detected binaries (right y axis) for the relative error in distance (blue solid line) and for the sky localisation error (red dashed line). The dashed vertical line marks our quality requirement $\sigma_{\rm d}/d < 0.3$, and the dotted horizontal line shows the fraction (number) of LISA detections that satisfies this requirement.}
         \label{fig:lisaperform}
\end{figure}
%%%%%%%%%%%%%%%%%%%%%%%%%%%%%

The distance can be found directly from the three GW observables $A, f$ and $\dot{f}$ by inverting eq.~\eqref{eqn:GWamp}:
\begin{equation}
d_{\rm GW} = \frac{5 c}{96\pi^2}\frac{\dot{f}}{f^3 A}.
\end{equation}
We compute the respective error as
\begin{equation} \label{eqn:sigmad}
\frac{\sigma_{\rm GW}}{d_{\rm GW}} \simeq  \left[\left(\frac{\sigma_{A}}{A}\right)^2 + \left(\frac{3\sigma_f}{f}\right)^2 + \left(\frac{\sigma_{\dot{f}}}{\dot{f}}\right)^2 \right]^{1/2},
\end{equation}
where ${\sigma_A}/{A}, {\sigma_f}/{f}$ and ${\sigma_{\dot{f}}}/{\dot{f}}$ are the diagonal elements of the covariance matrix provided by the MLDC pipeline (see Appendix~\ref{app:FIM} for a more detailed description). We verify that the terms containing correlation coefficients are at most of the order of 1\%, and we thus neglect them in eq.~\eqref{eqn:sigmad}.

The cumulative distribution (and total number) of the relative errors of the distance is represented in Fig.~\ref{fig:lisaperform}. Out of $2.6 \times 10^4$ binaries individually resolved by LISA only $30\%$ of the catalogue has relative distance errors of less than $30\%$, which nevertheless provides a sample of $7.8\times 10^3$ DWDs.
In particular, a subsample of $\sim 100$ DWDs ($0.4\%$ of all resolved binaries) has relative errors on the distance of less than $1\%$. These sources have  high frequencies ($>3\,$mHz) and high SNR ($>100$), and are located between 1 and 13$\,$kpc from the Sun. This remarkable precision is due to the fact that GW SNRs decrease much more slowly with distance compared to EM observations, and it is at the heart of the unique ability of the LISA mission to study the Milky Way's structure. The red solid line in Fig. \ref{fig:lisaperform} represents the sky localisation error, $\Delta \Omega = 2\pi \sigma_{\theta} \sigma_{\phi} \sqrt{1-\rho^2_{\theta \phi}}$ where $\rho_{\theta \phi}$ is the correlation coefficient between $\theta$ and $\phi$ \citep[e.g][]{Lang2008}, and shows that about half of all DWDs can be located to within  better than $10\,$deg$^2$ on the sky, with a maximum error in the whole sample of $\sim 100\,$deg$^2$.

\subsection{Distances from parallaxes}

%%%%%%%%%%%%%%%%%%%%%%%%%%%%%
\begin{figure}
        \centering
	 \includegraphics[width=0.46\textwidth]{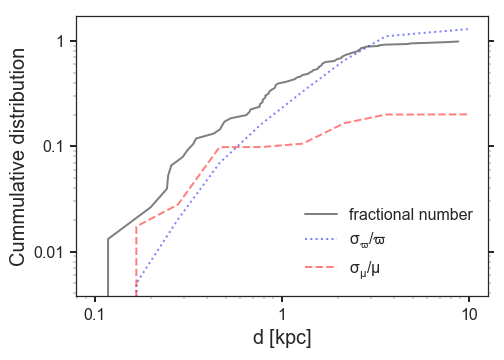} 
         \caption{Cumulative distribution of the LISA EM counterparts detected either by {\sl Gaia} or the LSST (grey solid line), and their median relative error in parallax (blue dotted line) and proper motion (red dashed line) as a function on the distance from us. For those DWDs that are detected by both {\sl Gaia} and the LSST we select the measurement with smaller uncertainty.}
    \label{fig:gaiaLSSTpermorm}
\end{figure}
%%%%%%%%%%%%%%%%%%%%%%%%%%%%%

To simulate the measurement of the parallax $\varpi$ for each optically detected DWD in our catalogue, we draw $\varpi$ from a Gaussian distribution centred on $1/d$ and with standard deviation $\sigma_{\varpi}$.
The {\sl Gaia} end-of-mission parallax error $\sigma_{\varpi}$ is given by \citep{Gaia}
\begin{equation}
\begin{aligned}
\sigma_{\varpi} = \Pi (-1.631 + 680.766z + 32.732z^2)^{1/2} & \times\\
\left[0.986 + (1 - 0.986) · (V-I) \right],
\end{aligned}
\end{equation}
where $z = \max \left[100.4 (12.09 - 15), 100.4(G - 15) \right]$, $V-I$ is the colour of the object in the Johnson-Cousins system, and $\Pi$ is a numerical factor that takes into account the Ecliptic latitude of the source and the number of transits of the satellite at that latitude\footnote{Tabulated values for $\Pi$ can be found at:

\url{https://www.cosmos.esa.int/web/Gaia/table-2-with-ascii} }.
To transform the colours of DWDs in our mock catalogue from the Sloan {\it ugriz} to the Johnson-Cousins UBVRI system, we use the empirical colour transformations from \citet{Jordi2006}.
We also calculate the end-of-mission errors on the proper motion ($\sigma_{\mu}$), which can be obtained by rescaling $\sigma_{\varpi}$ by a factor 0.526 \citep{Gaia}.
Note that we use the end-of-mission errors. To rescale the errors for a different observation time one needs to multiply $\sigma_{\varpi}$ by $(T_{\rm tot}/T_{\rm obs})^{0.5}$, where $T_{\rm tot}$ is the total {\sl Gaia} mission mission life time and $T_{\rm obs}$ is the effective observation time, both expressed in month \citep{Gaia2018}. For example, for the second {\sl Gaia} data release this factor is $\sim (60/21)^{0.5}$. For proper motion errors the scaling factor is $(T_{\rm tot}/T_{\rm obs})^{1.5}$.

We estimate the accuracy of the LSST astrometric measurements by interpolating Table 3.3 of \citet{LSST}. In the following, for the EM counterparts that can be detected by both {\sl Gaia} and LSST, we utilise the measurement of the parallax and proper motion with the smaller error.

\begin{figure}
        \centering
	 \includegraphics[width=0.45\textwidth]{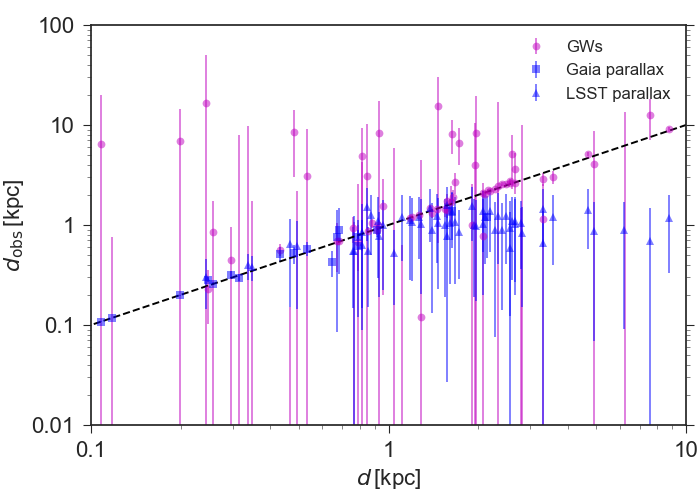} 
	 \includegraphics[width=0.45\textwidth]{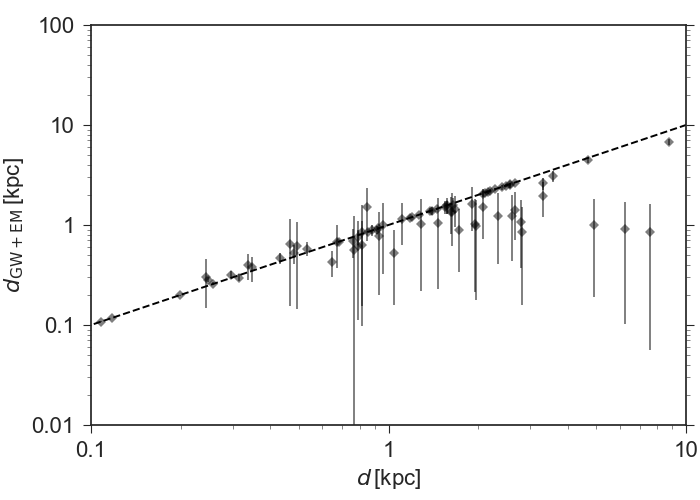} 
         \caption{In the top panel: observed distance as a function of the true distance to the DWD, $d$. We indicate with $d_{\rm obs}$ the distance estimated either from GWs (in magenta) or from parallax (in blue). We denote distances estimated respectively from {\sl Gaia} and LSST measurements with triangles and squares.  The dashed line shows where $d_{\rm obs} = d$. In the bottom panel:  distance estimates obtained by combining GW and EM measurements through Bayes theorem.} \label{fig:dest}
\end{figure}

In Fig. \ref{fig:gaiaLSSTpermorm}, we represent the cumulative distribution of the LISA EM counterparts (in grey), and that of their median relative error in parallax (in blue) and proper motion (in red) as a function of distance.
For binaries at $d<1\,$kpc the expected relative error in parallax is $<20\%$. These binaries constitute $30\%$ of the EM catalogue and  consists mainly of {\sl Gaia} measurements (see Fig.\ref{fig:dest}). Beyond $1-2\,$kpc all measurements are provided by the LSST. Although the median relative errors in parallax are larger, the LSST data is crucial in providing EM measurements out to $10\,$kpc. Forecasting the proper motion measurements, we show that the relative errors will be $<20\%$ at all distances.

Different authors have stressed that to correctly estimate distances from  parallaxes a probability-based inference approach is necessary \citep[e.g.][for {\sl Gaia} measurements]{BailerJones2015,Astraatmadja2016,BailerJones2018,Luri2018}.
Essentially, because the measurement of $\varpi$ is affected by  uncertainties, one can only infer the distance in a probabilistic sense by making an assumption on the true distribution of DWDs in space (the prior distribution).
Using  Bayes' theorem, the posterior probability density of the possible values of $d_{\rm EM}$ can be expressed as
\begin{equation} \label{eqn:bayes}
P(d_{\rm EM}|\varpi,\sigma_{\varpi})= \frac{1}{Z} P(\varpi|d_{\rm EM},\sigma_{\varpi}) P(d_{\rm EM}),
\end{equation}
where $Z$ is a normalisation constant, $P(\varpi|d_{\rm EM},\sigma_{\varpi})$ is the likelihood that describes the noise model of the instrument and $P(d_{\rm EM})$ is the prior.
We assume that the likelihood is Gaussian \citep[e.g.][]{Luri2018}.
For measurements with relative errors on parallax $\sigma_{\varpi}/\varpi\lesssim 0.2$, the distance estimates
are mainly independent of the choice of the prior. 
However, for larger relative errors the quality of the estimates depends on how well the prior describes the true distribution of distances of the observed sources.
In our sample we expect the choice of the prior to become crucial at $d>1\,$kpc.
For this work we adopt a simple exponentially decreasing volume density prior, described by only one parameter $L$, the scale length.
In this paper we assume $L=400\,$pc as in \citet{Kupfer2018}, and we fine-tuned this value by using our mock population to derive distances for LISA verification binaries using parallax measurements from the {\sl Gaia} Data release 2. 
We associate the most probable value of $d_{\rm EM}$ with the mode of the posterior distribution, because 
we expect this distribution to be highly asymmetric \citep[e.g.][]{BailerJones2015}.
Finally, we compute the errors as $\sigma_{\rm EM} = (d_{95}-d_5)/2s$, where $d_{95}$ and $d_5$ are the boundaries of the 90\% credible interval of the posterior distribution and $s=1.645$\footnote{ $s$ is the ratio of the 90\% to 68.3\% credible intervals for a Gaussian distribution.} \citep{BailerJones2015}.
The result is represented  in blue in the top panel of Fig. \ref{fig:dest}.
It is evident that distances inferred from parallaxes follow the dashed line $d_{\rm obs}=d$ up to $\sim 1-2\,$kpc, while beyond that the estimated distances systematically start to underestimate the true values.
This is due to the large parallax errors combined with our choice for the prior.
However, for these binaries more precise distances can be derived using additional information from GWs.

%%%%%%%%%%%%%%%%%%%%%%%%%%%%%%%%%%%
%%%%%%%%%%%%%%%%%%%%%%%%%%%%%%%%%%%%

\subsection{Combining GW and EM measurements} \label{sec:distEMGW}

\begin{figure}
        \centering
	 \includegraphics[width=0.45\textwidth]{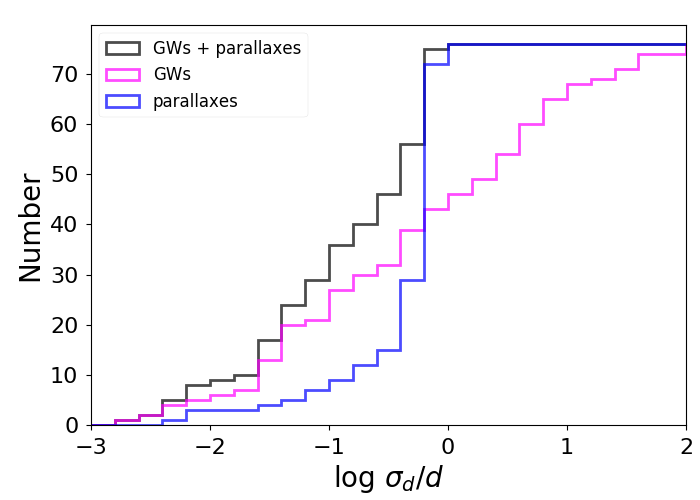} 
         \caption{The distribution of relative errors on the distance, estimated from GW observations (magenta), from optical observations (blue) and from the combination of the two measurements  (hatched).} \label{fig:em+gw}
\end{figure}

For DWDs with EM counterparts, we can use the additional information from EM observations to improve GW estimates.
Again, this can be done by using Bayes' theorem.
We model the GW posterior distribution for the distance as a Gaussian centred on the distance inferred from GWs, $d_{\rm GW}$, with a standard deviation equal to $\sigma_{\rm GW}$ (computed from the FIM $\Gamma$ as described in Sect. 3.1).
Likewise, we model the EM posteriors as a Gaussian centred on the distance inferred from the parallax, $d_{\rm EM}$, with a standard deviation equal to the corresponding error $\sigma_{\rm EM}$.
The joint posterior distribution is given by the product of these two Gaussian distributions. This can be understood from Bayes' theorem, by noting that the GW and EM observations are independent, and by using the GW posteriors as priors for the EM inference (or vice versa). The resulting distribution is  again  Gaussian with mean equal to the sum of the individual means weighted by their standard deviations,
\begin{equation}
d_{\rm GW + EM} = \frac{d_{\rm GW}\sigma^2_{\rm EM} + d_{\rm EM} \sigma^2_{\rm GW}}{\sigma^2_{\rm EM} + \sigma^2_{\rm GW}},
\end{equation}
and a standard deviation equal to twice the harmonic mean of the individual standard deviations,
\begin{equation}
\sigma_{\rm GW + EM} = \sqrt{ \frac{\sigma^2_{\rm EM} \sigma^2_{\rm GW}}{\sigma^2_{\rm EM} + \sigma^2_{\rm GW}}}.
\end{equation}
The result is represented in the bottom panel of Fig.~\ref{fig:dest}.
Comparing the top and bottom panels, it is evident that with this procedure we essentially select the best of the two measurements. Moreover, we also reduce the uncertainties compared to just selecting the more precise of the EM or GW measurements individually.
Indeed, in Fig. \ref{fig:em+gw} we show that by combining EM and GW data one can significantly improve the fractional errors on the distance, thus making it possible to use joint GW and EM detections to study Galactic kinematics, as we show in Section \ref{sec:5}.

%%%%%%%%%%%%%%%%%%%%%%%%%%%%%%%%%%%
%%%%%%%%%%%%%%%%%%%%%%%%%%%%%%%%%%%%

%%%%%%%%%%%%%%%%%%%%%%%%%%%%%%%%%%%%%%%%%%%

\begin{figure}
        \centering
	 \includegraphics[width=0.45\textwidth]{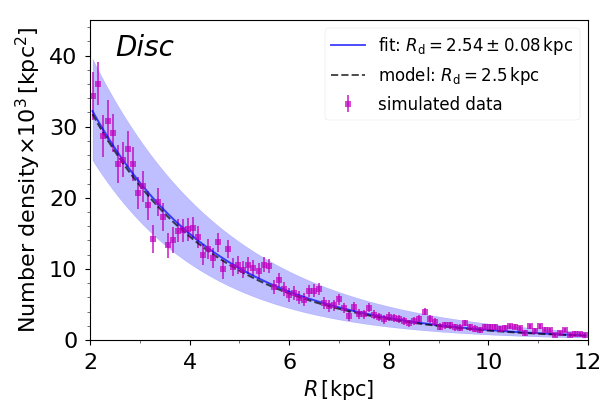} 
	 \includegraphics[width=0.45\textwidth]{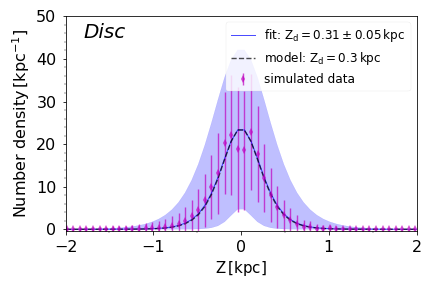}
	 \includegraphics[width=0.45\textwidth]{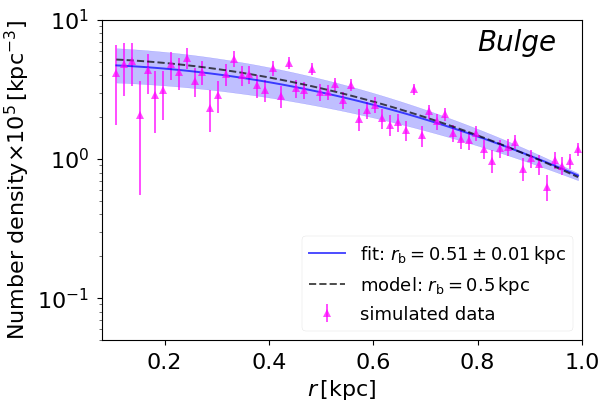}
         \caption{Number density profiles for the DWDs detected by LISA as a function of cylindrical radius $R$ (top panel), height above the Galactic plane $Z$ (middle panel), and spherical radius $r$ from the Galactic centre (bottom panel). Magenta points represent one of $10^5$ realisation of the LISA observations that we performed to compute the error bars. The blue solid line shows the best fit model and the blue shaded area shows its $3\sigma$ uncertainty region. The dashed grey line shows the true number density.}
       \label{fig:8}
\end{figure}

\section{Radial and vertical density profiles of LISA detections} \label{sec:4}

The distance and the sky localisation from LISA measurements allow one to construct density maps of DWDs in the Galaxy. Figure \ref{fig:3} suggests that LISA has the potential to reconstruct the density profiles of both the disc and bulge components and derive their scale lengths. In this section we quantify how well we can recover the scale parameters of the Milky Way using DWDs.

We define a Cartesian Galactocentric reference frame $(X,Y,Z)$ such that the Galactic disc lies on the $(X,Y)$ plane, and the Sun lies on the positive $Y$-axis in the Galactic plane (see also Fig. \ref{fig:sketch}).
In this reference frame, the position of an object with Galactic coordinates $(l,b)$ at a distance $d$ from the Sun is defined by the set of coordinates:
\begin{equation} \label{eqn:coord}
\begin{aligned}
X =& d \sin{l}\cos{b}, \\
Y =& R_{\odot} - d\cos{l}\cos{b}, \\
Z =& d \sin{b}.
\end{aligned}
\end{equation}
In addition, we define a cylindrical coordinate system about the Galactic centre as
\begin{equation} 
\begin{aligned}
R =& \sqrt{X^2+Y^2}, \\
\theta =& \arctan \frac{Y}{X}, \\
Z =& d \sin{b}.
\end{aligned}
\end{equation}
We select the subsample of LISA detections with relative error in distance $<30\%$.
This leaves us with $\sim 8 \times 10^3$ DWDs ($30\%$ of all the binaries detected by LISA).
To compute the radial density profile,  we first derive $10^5$ realisations of the 3D binary positions in the Galaxy by randomly drawing $l,b$ and $d$ from Gaussian distributions\footnote{We consider the three Gaussian distributions independent because the correlation coefficients between $d, \theta$ and $\phi$ are negligible: $\rho_{\rm d \theta}, \rho_{\rm d \phi } \le 0.1$ and $\rho_{\theta \phi} < 0.3$ in our catalogue.} centred on their true values and with standard deviations computed in Sect.~\ref{sec:distances}. 
For each realisation we compute the cylindrical Galactocentric distance, $R$, and we select sources with $2 \le R \le 12\,$kpc. 
The lower limit of the interval in $R$ is motivated by the number density maps  represented in Fig.~\ref{fig:3}, which show a spherical central population in the inner $\sim 2\,$kpc, which we identify with the bulge.
The upper limit is motivated by the poor statistics at $R > 12\,$kpc, as can be seen in Fig.~\ref{fig:3}.
Next, we count the number of DWDs in cylindrical shells of width $dR = 0.125\,$kpc, dividing by the shell volume and accounting for the bias (Sect. \ref{app:bias}).
We compute the error on the number density in each bin as the standard deviation over different realisations.
We represent one of the data realizations by the square symbols in the upper panel of Fig. \ref{fig:8} (upper panel). 
We fit the scale radius $R_{\rm d}$ and the normalisation with {\sc PyMC3}\footnote{{\sc PyMC3} is an open source python package for Bayesian statistical modelling and probabilistic machine learning \citep{pymc3}.}, using an exponential profile (eq. \eqref{eqn:disc}).
The blue solid curve in the top panel of Fig. \ref{fig:8} shows the best fit model, and the coloured area shows its $3\sigma$ interval.
Our best fit value for the disc scale radius is $R_{\rm d} = 2.54 \pm 0.08\,$kpc, in agreement with the fiducial value of $2.5\,$kpc that we use to generate the Galaxy.
Thus, LISA can recover the disc scale radius with $\sim 3\%$ precision.

To study the vertical distribution of DWDs in the disc, we select binaries  with  $2 \le R \le 12\,$kpc. First, we bin them in concentric cylindrical rings with a step of $0.125\,$kpc in the radial direction and $0.05\,$kpc in the vertical direction. Next, we divide the bin counts by the bin volume $2\pi R dR dZ$.
In each radial bin, we model the number density with a $\sech^2 (Z/Z_{\rm d})$ function and fit $Z_{\rm d}$ to test whether the scale height is constant with $R$ or the vertical distribution of DWDs has a more complex structure.
We find a constant behaviour and therefore we decide to increase the statistics by computing the average value of $Z_{\rm d}$ and its error on a stacked radial profile. 
In this way, we find $Z_{\rm d} = 0.31 \pm 0.05\,$kpc, which is consistent with the fiducial value  of $0.3\,$kpc.

Finally, to estimate the scale radius of the bulge we select DWDs in the inner $1.2\,$kpc to avoid disc contamination.
Again, we compute $10^5$ realisation of the binary positions in the Galaxy by randomly drawing $l,b$ and $d$ for each source.
For each realisation, we  estimate the number density profile by counting DWDs in spherical shells with radius $r = \sqrt{X^2+Y^2+Z^2}$ and $dr = 15\,$pc, dividing this number by the shell volume and correcting for the bias  (Sect. \ref{app:bias}).
Finally, we estimate the error in each bin as the standard deviation over all the realisations.
The result is given by the magenta triangles in the bottom panel of Fig. \ref{fig:8}.
To fit the scale radius of the bulge, we use  eq.~\eqref{eqn:bulge} as the model distribution, and we obtain $r_{\rm b}=0.51 \pm 0.013\,$kpc.
Again, this result is in excellent agreement with the fiducial value of $0.5\,$kpc (see Tab. \ref{tab:1}).

\subsection{Model comparison for the disc radial density profile }

%%%%%%%%%%%%%%%%%%%%%%%%%%%%%%%%%%%%%%%%%%%%%
\begin{figure}
        \centering
	 \includegraphics[width=0.45\textwidth]{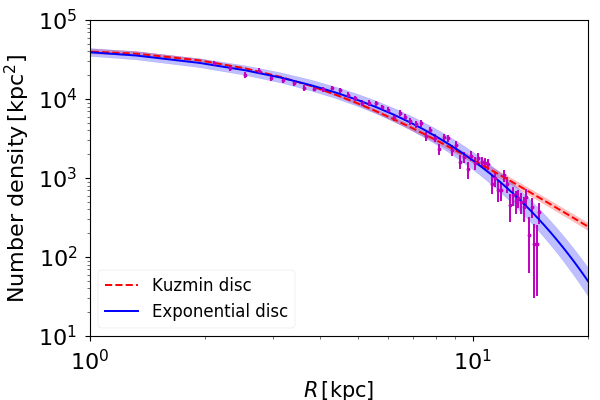} 
	 \includegraphics[width=0.45\textwidth]{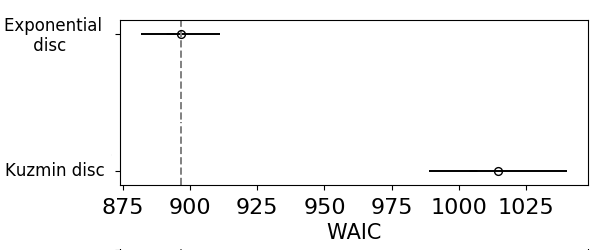} 
         \caption{The top panel shows number density profiles for the exponential and Kuzmin disc models as a function of $R$: magenta points represent simulated data like in Fig. \ref{fig:8}, blue solid and red dashed lines show the best fits for the exponential disc and Kuzmin disc models respectively, and the shaded areas define $3\sigma$ uncertainties. The bottom panel shows a comparison between the two models in terms of the WAIC criterion: empty circles represent the WAIC value,  and the black error bars are the associated errors computed using  {\sc PyMC}3; the vertical dashed line marks the preferred model.}
       \label{fig:9}
\end{figure}
%%%%%%%%%%%%%%%%%%%%%%%%%%%%%%%%%%%%%%%%%%%%%

Heretofore, we have tested how well the simulated GW data trace the underlying density distribution (i.e. the true model).
In this Section we assess whether the simulated data allow us to discriminate between the true disc surface density distribution and a model with a different functional form.

We consider a Kuzmin disc \citep[Kuzmin 1956,][]{ Toomre1963}, whose surface density distribution scales as a power law:
\begin{equation} \label{eqn:kuzmin}
\Sigma_{\rm K }(R) = \frac{M_{\rm d}}{2\pi  (R^2 + R_{\rm K}^2)^{3/2} }  \ \  {\text M}_{\odot}\, {\text{kpc}}^{-2},
\end{equation}
where $M_{\rm d}$ is the mass of the disc and $R_{\rm K}$ is the model's radial scale parameter.
Unlike our ``true'' disc model, whose surface density profile decays exponentially with $R$, eq.~\eqref{eqn:kuzmin} yields $\Sigma_{\rm K}(R) \propto R^{-3}$ at large $R$.
Thus, we expect the two models to differ significantly at least at large $R$. 

We fit the simulated data with eq.~\eqref{eqn:kuzmin},
and obtain $ R_{\rm K} = 3.86 \pm 0.09\,$kpc. We  show  in Fig. \ref{fig:9} a comparison between the best fit to the Kuzmin model (in red) and the best fit to the exponential disc model (in blue).
This figure reveals that the two models are indistinguishable inside the Solar Galactocentric radius, and start differing beyond that radius.
Therefore, to distinguish between these two models  data far out in the disc are needed,which GW detections can provide (magenta circles in Fig. \ref{fig:9})

We therefore compare the two models using the Widely-applicable Information Criterion (WAIC), which provides a fit measure for Bayesian models and which can be applied when the parameter estimation is done using numerical techniques \citep{Watanabe2010}.
The WAIC is defined as
\begin{equation} \label{eqn:waic}
WAIC = -2\  (LPPD - \bar{P}), 
\end{equation}
where $LPPD$ is the log posterior predictive density, and $\bar{P}$ is an estimate of the effective number of free parameters in the model, which can be interpreted as a penalty term adjusting for overfitting\footnote{A higher value of $\bar{P}$ indicates that the model is the more ``flexible'' of the two at fitting the data.}.
By definition, lower values of the WAIC indicate a better fit, i.e the WAIC measures the ``poorness'' of the fit.
We compute the WAIC ($\bar{P}$) with {\sc PyMC3}, obtaining 895 (2.12) and 1017 (4.6) for the exponential and Kuzmin disc models respectively (see bottom panel of Fig.~\ref{fig:9}).
There is no set threshold for the difference in WAIC, but typically a difference of 10 or more suggests that the model with higher WAIC is likely to perform worse. 
Thus, in our case, the Kuzmin disc model is more ``flexible'' with respect to the data, but its predictive power is worse than the exponential disc model. Furtermore, the error on the WAIC (the expected predictive error) is also larger for the Kuzmin disc (Fig.~\ref{fig:9}).
These factors reveal a preference for the correct exponential disc model.

%%%%%%%%%%%%%%%%%%%%%%%%%%%%%%%%%%%%%%%%%%%%%%%%%%%%%%%%%%%%%%%
%%%%%%%%%%%%%%%%%%%%%%%%%%%%%%%%%%%%%%%%%%%%%%%%%%%%%%%%%%%%%%%
\section{Kinematics of DWDs} \label{sec:5}

In the previous Section we have shown that one can recover the shape of the baryonic components of the Galaxy from GW observations alone, but EM counterparts are required to study the dynamics of the Galaxy.
Around $80$ DWD EM counterparts to LISA detections can be observed with {\sl Gaia} and the LSST through their eclipses (Sect. \ref{sec:EMdetections}).
We estimate that both {\sl Gaia} and the LSST will deliver proper motions with relative precision $<20\%$ for these binaries. 
However, it will be hard to have 3D velocities without a spectroscopic follow-up of these sources.
DWDs are too faint to measure their radial velocities with the Radial Velocity Spectrometer (RVS) on board of the {\sl Gaia} satellite and, moreover, they are typically featureless in the RVS wavelength range \citep{Carrasco2014}. 
Nonetheless, the rotation speed of DWD EM counterparts around the Galaxy can be computed  from proper motions alone \citep[e.g.,][]{Sofue2017}.
In this section we describe how we model DWD velocities, and we derive the rotation curve for our mock Galaxy using distances estimated from GW observations as well as proper motions simulating {\sl Gaia} and the LSST observations.

\subsection{Kinematic model}

\begin{figure}
        \centering
	 \includegraphics[width=0.35\textwidth]{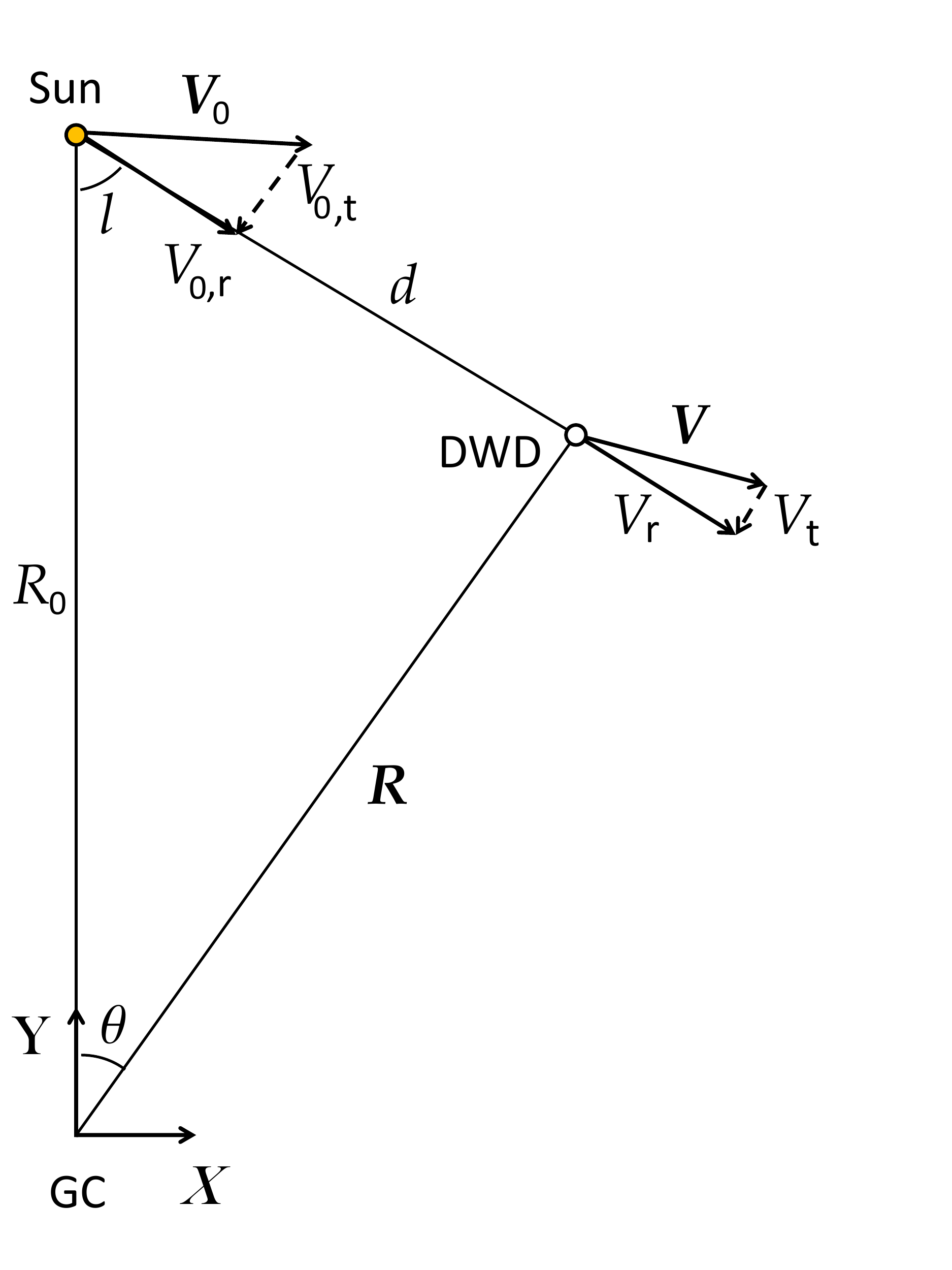} 
         \caption{Kinematic model for DWDs. GC stands for Galactic centre.}
       \label{fig:sketch}
\end{figure}

Figure \ref{fig:sketch} sketches the geometry of the problem:
a DWD at a distance $d$ from the Sun and at Galactic latitude $l$ is moving along a circular orbit in the Galactic plane, with Galactocentric radius $R$.
In the Cartesian coordinate system  defined by the coordinate transformation of eq.~\eqref{eqn:coord}, the position vector of the binary can be expressed as
\begin{equation}
{\mathbf R} =  \begin{pmatrix} R \sin \theta \\ R \cos \theta \end{pmatrix} =  \begin{pmatrix} d\sin l \\ R_0 - d\cos l \end{pmatrix}\,,
\end{equation} 
where $\theta$ is the angle between the Sun and the DWD as seen from the Galactic centre.
By equating the two expressions for the components of ${\mathbf R}$, one obtains $\sin \theta = d \sin l /R$ and  $\cos \theta = (R_0-d\cos l)/R$.
Thus, we can write the azimuthal velocity as
\begin{equation}
{\mathbf V} = V(R) \begin{pmatrix} \cos \theta \\ - \sin \theta \end{pmatrix} = V(R) \begin{pmatrix} \frac{R_0}{R} - \frac{d}{R} \cos l \\ - \frac{d}{R} \sin l \end{pmatrix},
\end{equation} 
In practice, we assign a value   of $V(R)$ to a source by randomly drawing from a Gaussian centred on the value given by the rotation curve at that $R$ and dispersion given by eq. \eqref{eqn:sigma_vphi}. If we neglect the peculiar motion of the Sun and assume that its velocity in the Galactic plane is ${\mathbf V}_{\odot} = (V_0, 0)$, we can write the relative velocity between the DWD and the Sun as
\begin{equation}
\Delta {\mathbf V} = \mathbf{V - V}_{0}= \begin{pmatrix} R_0(\Omega(R) - \Omega_0) - \Omega(R) d \cos l  - \Omega(R) d\\ - \Omega(R) d \sin l \end{pmatrix} ,
\end{equation} 
where $\Omega(R)=V(R)/R$ and $\Omega_0=V_0/R_0$ are the angular velocities of the DWD and of the Sun, respectively.
Then, the tangential component can be found by projecting $\Delta {\mathbf V}$ along the line of sight and along the direction perpendicular to it:
\begin{equation}
V_{\rm t} = \Delta{\mathbf V} \begin{pmatrix} \cos l \\ \sin l \end{pmatrix} = \left[ \Omega(R)-\Omega_0 \right] R_0 \cos l - \Omega(R) d.
\end{equation} 
The proper motions of DWDs can be estimated as
\begin{equation}
\mu = \frac{V_{\rm t}}{4.74 \ d} \qquad  \text{arcsec yr}^{-1},
\end{equation}
where $d$ is in pc and $V_{\rm t}$ is in km/s.

%%%%%%%%%%%%%%%%%%%%%%%%%%%%%%%%%%%

\begin{figure}
        \centering
	 \includegraphics[width=0.5\textwidth]{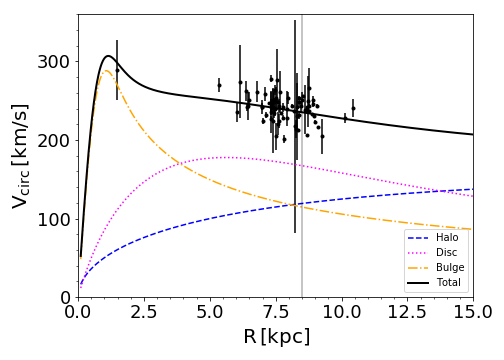} 
         \caption{Rotation speed of DWDs with EM counterpart computed according to eq.~\eqref{eqn:vrot_obs}. The black solid curve shows the model's rotation curve. Coloured lines represent the contributions of different Galactic components to the total rotation curve: the colour coding is the same as in Fig.~\ref{fig:MWrotcurve}. The vertical line marks the position of the Sun.}
       \label{fig:rot_curve}
\end{figure}
%%%%%%%%%%%%%%%%%%%%%%%%%%%%%%%%%%%%

To simulate {\sl Gaia} and LSST measurements of DWD proper motions, we assign  an observed proper motion $\mu_{\rm obs}$ to a source by sampling from a Gaussian centred on $\mu$ with an error $\sigma_{\mu}$ given by the instrument response (see Sect. \ref{sec:EMdetections}). Similarly we sample the observed distances from a Gaussian centred on $d_{\rm GW+EM}$ with an error $\sigma_{\rm GW+EM}$ (see Sect. \ref{sec:distEMGW}). To compute the observed rotation speed we combine the simulated measurements according to
\begin{equation} \label{eqn:vrot_obs}
V_{\rm obs}(R) = - \frac{R}{d_{\rm obs} - R_0 \cos l} \left(4.74 \mu_{\rm obs} d_{\rm obs} + V_0 \cos l \right) \qquad \text{km s}^{-1}.
\end{equation}
For each DWD, we calculate $V_{\rm obs}(R)$ for $10^5$ independent realizations of $\mu_{\rm obs}$ and $d_{\rm obs}$, and we assign an observed velocity and measurement error equal respectively to the mean and the standard deviation of the resulting distribution of $V_{\rm obs}(R)$.  
The result is represented in Fig. \ref{fig:rot_curve}.
Because {\sl Gaia} and LSST can probe only relatively close distances, the rotation curve derived here can provide information only on the local Galactic properties.
However, the one observation point that we have close to the Galactic centre provides good constraints on the parameters describing the bulge component, as we show in the following.

\subsection{Doppler effect due to motion in the Galaxy}

In this Section, we calculate the line of sight projection of the velocity, $V_{\rm r}$,  which for DWDs will not be observed by Gaia and/or LSST, but which will directly influence the GW observables, as we explain below.

The motion of the stars in the Galaxy introduces a Doppler shift in the GW frequency, so that the observed frequency is
\begin{equation} \label{eqn:Doppler_f}
f_{\rm obs}  \approx \frac{ f}{1 + \frac{V_{\rm r}}{c}},
\end{equation}
where $V_{\rm r}$ can be computed by projecting ${\mathbf V}$ along the line of sight, i.e.
\begin{equation}
V_{\rm r} = \Delta{\mathbf V} \begin{pmatrix} \sin l \\ -\cos l \end{pmatrix} = \left[ \Omega(R)-\Omega_0 \right] R_0 \sin l.
\end{equation}
The relation between  time intervals at the detector and at the source is
\begin{equation} \label{eqn:Doppler_t}
dt_{\rm obs}  \approx \left(1 + \frac{V_{\rm r}}{c} \right) dt.
\end{equation}
By deriving eq.~\eqref{eqn:Doppler_f} with respect of time and using ~\eqref{eqn:Doppler_t} to express the result in terms of the observed frequency we obtain
\begin{equation} \label{eqn:doppler}
	\dot{f}_{\rm obs} = \frac{96}{5} \pi^{8/3} \left[ \frac{G{\cal M} (1+V_{\rm r}/c)}{c^3} \right]^{5/3} f_{\rm obs}^{11/3} + \frac{\dot{V}_r}{c}f_{\rm obs}
\end{equation}
and the GW amplitude as
\begin{equation} \label{eqn:amp_doppler}
A = \frac{5}{96\pi^2}\frac{\dot{f}_{\rm obs}}{f_{\rm obs}^3d(1+V_{\rm r}/c)}.
\end{equation}
There are two additional terms in eq.~\eqref{eqn:doppler} compared to the original  eq.~\eqref{eqn:GWfdot}: the Doppler term containing $V_{\rm r}/c$ and the acceleration term  $\dot{f}_{\rm acc} = f_{\rm obs}\dot{V}_r/c$.

First, we focus on the Doppler term.
In the first term of eq.~\eqref{eqn:doppler} we can replace the chirp mass with the Doppler-shifted chirp mass ${\cal M} (1+V_{\rm r}/c)$. Similarly, the Euclidean distance $d$ in eq.~\eqref{eqn:amp_doppler} can be replaced with the luminosity distance $d (1+V_{\rm r}/c)$.\footnote{Note that in the presence of a Doppler shift, the luminosity distance -- i.e. the ratio $L/(4 \pi F)$, $L$ being the intrinsic source luminosity and $F$ being the energy flux at the detector -- differs from the Euclidean distance $d$, because energies are red-(blue-) shifted and times are dilated (contracted).} This is similar to what happens for cosmological sources, for which the chirp mass gets ``redshifted'' (i.e. multiplied by a factor $1+z$, $z$ being the redshift), the frequency at the source gets
replaced by the detector-frame one, and the co-moving distance is replaced by the luminosity distance.
The radial velocities of DWDs as seen from the Sun are expected to be from a few to a few tenths km/s, meaning that $V_{\rm r}/c \sim 10^{-5} - 10^{-4}$. 

Next, we estimate the acceleration term $\dot{f}_{\rm acc}$. Assuming that the total  velocity of a DWD (relative to the observer) is constant, we can express $\dot{V}_r$ in terms of $V_{\rm t}$ as $\dot{V}_{\rm r} = V_{\rm t}^2/d$. 
For a DWD with a typical frequency of $1\,$mHz, tangential velocity of $10\,$km/s and distance of $1\,$kpc, we obtain  $\dot{f}_{\rm acc} \sim 10^{-23}\,$s$^{-2}$, meaning that the contribution of the acceleration term is $\dot{f}_{\rm acc}/\dot{f}_{\rm obs} \sim 10^{-5} - 10^{-4}$, thus comparable to the Doppler term.
The same applies to all periodic phenomena, and has long been known in classical astronomy as ``secular acceleration'' or ``Shklovsky effect'' in pulsar timing \citep[e.g.,][]{Shklovskii1970,paj95}.

In general, both the acceleration and the Doppler terms are too small to influence LISA's measurements.
However, \citet{Shah2014} have shown that for DWDs with high frequencies and high SNRs, $\sigma_{\dot{f}}/\dot{f}$ can be determined with accuracy up to $10^{-4}-10^{-5}$, i.e. of the same order of magnitude as the two effects discussed here.
Consequently, for these high frequency binaries the systematic errors on $\dot{f}$ (and thus on the distance) due to the motion in the Galaxy can be $\sim 10\%$.
We do not take this into account in the present work, but we suggest that when estimating parameters for high frequency binaries, the Doppler effect and the acceleration term due to the motion in the Galaxy can introduce non-negligible systematic errors.

\subsection{Rotation curve fitting}

Although our model is  simpler than more realistic representations of the Milky Way (we do not account e.g. for the spiral arms and the bar), as many as seven parameters are required to fully characterise its rotation curve: $M_{\rm b}, r_{\rm b}, M_{\rm d}, R_{\rm d}, Z_{\rm d}, \rho_{ \rm h}$ and $r_{\rm h}$. 
In general, the measurement of the rotation speed alone is not sufficient to derive all the parameters and to break the degeneracies between them. 
A well known degeneracy is that between disc and halo parameters, i.e. 
a smooth flat rotation curve, such as the one of the Milky Way, makes the transition from the disc dominated  to the  DM halo dominated regime very gentle. The measurement of the rotation speed of stars in the Galaxy provides the total enclosed mass at a given radius, but in general that is not enough to break the degeneracy between the mass and the scale radius of the DM halo and disc components.
Thus, a global rotation curve fitting requires strong prior assumptions on the scale lengths of the Galactic components.

To obtain the best set of parameters that reproduce our simulated rotation curve (Fig. \ref{fig:rot_curve}), we fix $r_{\rm b}, R_{\rm d}$ and $Z_{\rm d}$ to the values obtained by fitting the number density profiles of DWDs, and we fit the remaining parameters using {\sc PyMC3}.
We use as  proposal fitting model the rotation curve computed numerically with  {\sc galpynamics} according to eq.\eqref{eqn:rot_curve}, and we leave $\rho_{\rm h}, r_{\rm h}, M_{\rm d}$ and $M_{\rm b}$  as free parameters of the model.
For all four free parameters, we set flat uninformative priors in the following ranges: $M_{\rm d}$ and $ M_{\rm b}$ are searched between $(1 - 10) \times 10^{10}$M$_{\odot}$; $\rho_0$ and $r_{\rm h}$ between $(0.1 - 10) \times 10^7\,$M$_{\odot}$/kpc$^3$ and $10 - 30\,$kpc, respectively.
At each MCMC step we evaluate the value of the likelihood times the priors by computing the difference between our model and the simulated observations.
The final posterior probability distribution of the free parameters is represented in (Fig. \ref{fig:rot_curve}).
It shows that DWDs can recover the mass of the disc and bulge components, but not that of the DM halo.
This is because there is no data at $R > 11\,$kpc (Fig.~\ref{fig:3}), where the halo dominates the dynamics in our Milky Way model.
We estimate the mass of the disc to be $M_{\rm d} = 5.3^{+1.29}_{-1.71} \times 10^{10}\,$M$_{\odot}$ and the mass of the bulge to be $M_{\rm b} = 2.49^{+0.44}_{-0.42} \times 10^{10}\,$M$_{\odot}$, in good agreement with our fiducial values. 
Remarkably, our constraints on the bulge mass are extremely competitive with those derived from EM tracers \citep[see e.g.][]{Bland-Hawthorn2016}. 
The larger errors on the disc mass stem from our choice to leave the halo parameters unconstrained.

\begin{figure*}
        \centering
	 \includegraphics[width=0.9\textwidth]{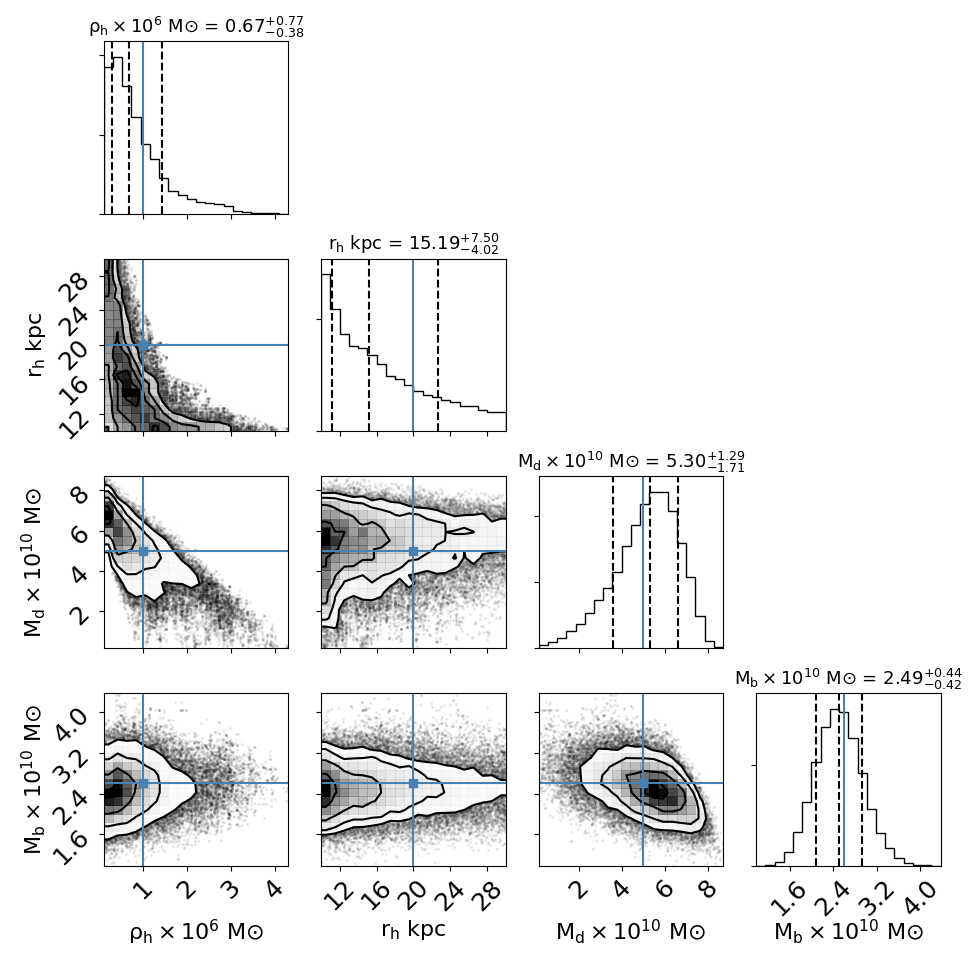} 
         \caption{The posterior probability distribution of the four free parameters of our rotation curve fitting model, $M_{\rm b},  M_{\rm d}, \rho_h$ and $r_{\rm h}$. Blue lines mark the true values listed in Tab. \ref{tab:1}}
       \label{fig:final}
\end{figure*}
\begin{figure*}
        \centering
	 \includegraphics[width=0.9\textwidth]{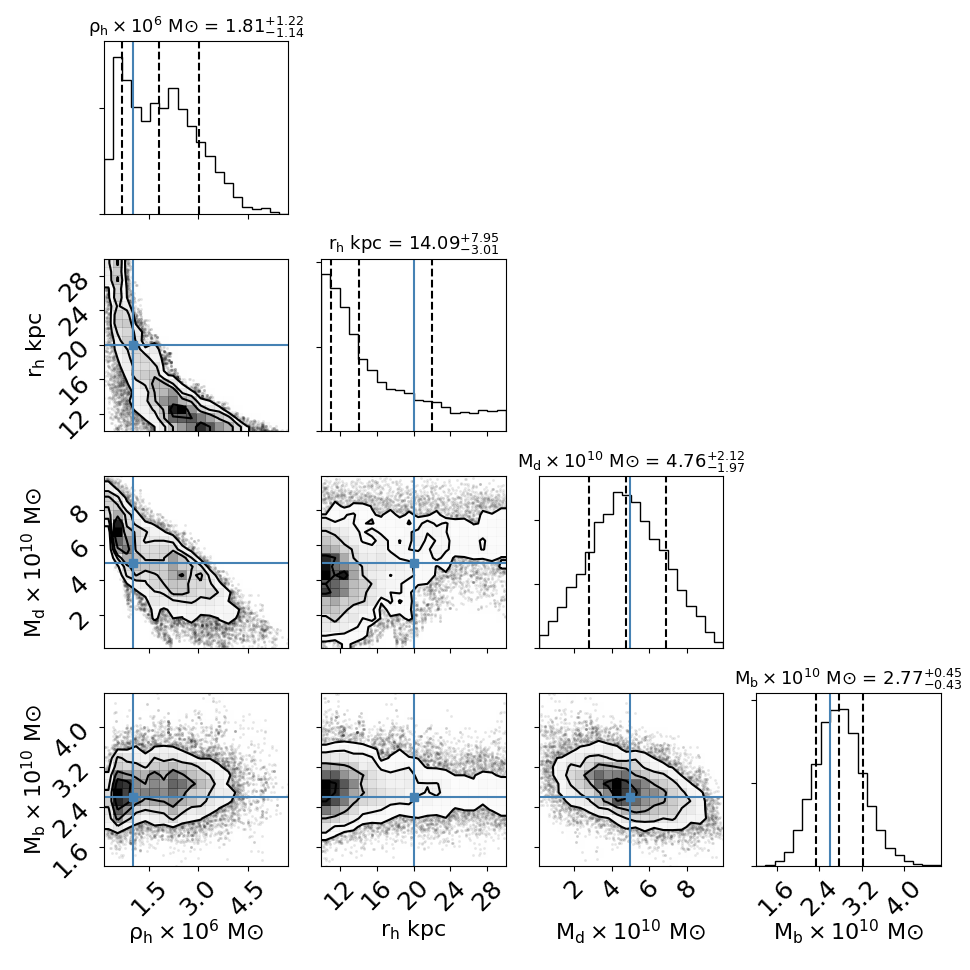} 
         \caption{The same as in Fig. \ref{fig:final}, but considering a Galaxy with a heavier DM halo ($M_{\rm h} = 10^{12}$M$_{\odot}$), so that at the Sun position the disc and the DM halo give comparable contributions to the total rotation curve. This ensures that our simulated observations sample the region of the Galaxy where DM is significant.}
       \label{fig:final1}
\end{figure*}

To test whether our method can provide better constraints on the DM halo component, we performed an additional simulation with a heavier DM halo, which gives a larger contribution to the total rotation speed  at the Sun position (where most of the data points lie). 
Specifically, we performed an additional simulation of DWDs kinematics (as described in Sect. \ref{sec:5}), in which we assign velocities to LISA optical counterparts using our fiducial Milky Way potential (with scale radius $r_{\rm h}=20\,$kpc), but with $\rho_{\rm h} =10^7\,$M$_\odot{\rm kpc}^{-3}$. This way the total mass of the DM halo is $M_{\rm h} \simeq   10^{12}\,$M$_\odot$, as found e.g. by \citet{Rossi2017}. By performing the same fitting procedure as above, we obtain the posterior probability density distributions for $\rho_{\rm h}, r_{\rm h}, M_{\rm d}$ and $M_{\rm b}$, represented in Fig. \ref{fig:final1}. Again, we obtain $M_{\rm b} = 2.77^{+045}_{-0.43} \times 10^{10}\,$M$_{\odot}$, which within a $1\sigma$ uncertainty recovers the true value of $2.5 \times 10^{10}\,$M$_{\odot}$. Although with large uncertainties, we can now recover also the true values of the DM halo parameters, $\rho_{\rm h}$ and $r_{\rm h}$. However, by comparing Figs. \ref{fig:final} and \ref{fig:final1}, it is evident that this degrades the uncertainty on the disc mass by a factor of $\sim 1.5$, highlighting the degeneracy between the disc and the halo components.
Thus, an improvement of this analysis should involve including additional information from DM halo tracers.

%%%%%%%%%%%%%%%%%%%%%%%%%%%%%%%%%%%%%%%%%%%%%%%%%%%%
\section{Conclusions} \label{sec:6}

In this study, we quantitatively investigate for the first time the prospects for tracing the baryonic mass of the Galaxy with a multi-messenger (GW+EM) data analysis using DWD binaries. 
The advantages over traditional tracers include the possibility of looking through the bulge, and beyond, thus allowing one to map both sides of the Galaxy using the same tracer.
We show that this unique property allows one to recover the scale radii of the baryonic components accurately and with  percent precision. The abundance of GW detections at large distances will also enable one to disentangle different disc stellar density profiles.
Finally, in synergy with optical data, GW measurements will provide competitive mass estimates for the bulge and stellar disc.

Our encouraging analysis, however, needs to be further tested against more realistic Milky Way potentials including, for example, spiral arms and other density asymmetries. One possible way to perform such a test is to use the matter distributions resulting from cosmological simulations of Milky Way like galaxies such as the Eris, APOSTLE and FIRE simulations \citep[][]{Guedes2011,saw16,hop18}.
Furthermore, we should also assess the impact of adding observations of AM CVn stars (ultra-compact accreting WDs), which although likely less numerous, may be seen at larger distances in the optical band due to their accretion luminosity.

Finally, our choice to use GW sources and their EM counterparts limits our ability to  constrain the DM halo component of Galaxy. This highlights the importance of a more precise knowledge of the DM halo to improve baryonic mass measurements. 
We therefore envisage that the full potential of our method can be unleashed when more stringent priors on the halo mass from DM tracers will be available after the full exploitation of {\sl Gaia} data \citep[e.g.][]{Posti2018, con18}.

%%%%%%%%%%%%%%%%%%%%%%%%%%%%%%%%%%%%%%%%%%%%%%%%%%%%%%%%%%%
\section*{Acknowledgements}
We thank G. Iorio, A. G. Brown, N. Tamanini, A. Petiteau, S. Babak and L.T. Maud for insightful comments.
VK would like to thank Dan Coe for  useful and concise explanation on Fisher Matrices detailed in \citet{Coe2009}.
This research made use of {\sc galpynamics, NumPy, SciPy, PyMC3} \citep{pymc3}, {\sc corner.py} \citep{corner} and {\sc PyGaia} python packages and matplotlib python library. 
This work was supported by NWO WARP Program, grant NWO 648.003004 APP-GW. This project has received funding from the European Union's Horizon 2020 research and innovation programme under the Marie Sklodowska-Curie grant agreement No 690904.
\addcontentsline{toc}{section}{Acknowledgements}

%%%%%%%%%%%%%%%%%%%%%%%%%%%%%%%%%%%%%%%%%%%%%%%%%%

%%%%%%%%%%%%%%%%%%%% REFERENCES %%%%%%%%%%%%%%%%%%

\bibliographystyle{mnras}
\bibliography{biblio}

%%%%%%%%%%%%%%%%%%%%%%%%%%%%%%%%%%%%%%%%%%%%%%%%%%

%%%%%%%%%%%%%%%%% APPENDICES %%%%%%%%%%%%%%%%%%%%%

\appendix

\section{} \label{app:FIM}

The measurement precision of the parameters describing the waveform can be forecast by computing the FIM, commonly denoted by $\Gamma$ \citep[e.g.][]{Cutler1998, Shah2012}.
The GW waveform produced by a DWD can be characterised by 9 parameters: $A, f, \dot{f}, \ddot{f}, \theta, \phi, \iota, \psi$ and $\phi_0$, thus  $\Gamma$ is a $9 \times 9$ matrix.
The components of $\Gamma$ can be computed as 
\begin{equation}
\Gamma_{ij} = \frac{2}{S_n(f)} \sum_{\alpha = I, II} \int_0^{T_{\rm obs}} dt \ \partial_i h(t) \partial_j h(t),
\end{equation}
where we assume that for a quasi-monochromatic source the noise power spectral density at the binary GW frequency, $S_n(f)$, is constant over the lifetime of the LISA mission  and $\alpha = I, II$ are the two independent two-arm detectors of the LISA current design \cite[e.g.][]{Cutler1998,Takahashi2002, Seto2002}. We adopt the noise power spectral density $S_n(f)$ from \citet{LISA2017}.  
The inverse of the FIM is the covariance matrix, $C$.
The diagonal elements of the covariance matrix represent squared $\sigma$ parameter uncertainties, while the off-diagonal elements give the covariances between parameters.
To compute the uncertainty on the distance ($\sigma_{\rm GW}$) we first marginalise over the parameters that do not enter the distance determination ($\ddot{f}, \theta, \phi,  \psi$ and $\phi_0$) by removing the corresponding rows and columns  from the covariance matrix. 
Next, we invert the resulting covariance matrix to obtain a 4x4 FIM in terms of $p = (f, A, \iota, \dot{f})$ only, and we compute the new FIM in terms of new parameters $p' =  (f, d, \iota, \dot{f})$:
\begin{equation}
\Gamma'_{mn} = \sum_{ij} \frac{\partial p_i}{\partial p'_m} \frac{\partial p_j}{\partial p'_n} \Gamma_{ij}.
\end{equation}
Finally, the second diagonal element of the inverse of $\Gamma'$ represents $\sigma^2_{\rm GW}$.
We confirm that the results obtained in this manner are equivalent, within $0.001\%$, to the approximate expression in eq.~\eqref{eqn:sigmad}.
Since eq.~\eqref{eqn:sigmad} does not account for the correlations between $A, f$ and $\dot{f}$, this excellent agreement must imply that these correlation terms are negligible. We have indeed verified that this is the case.
Note that in general $\sigma_{\rm GW}/{d_{\rm GW}} $ is small for  binaries with small $\sigma_{\dot{f}}/\dot{f}$, i.e. whose chirp is larger than the instrument resolution in frequency ($\dot{f} T_{\rm obs} > 1/T_{\rm obs}$). Thus, a precise distance measurement is typically more challenging for DWDs than for e.g. massive black holes, because the former  evolve gravitationally more slowly (eq.\eqref{eqn:tau}) in the observation window and because they have much smaller masses. However, within the Galaxy, the abundance of DWD binaries is such that we can collect a sizeable sample with good distance determinations. 

\section{LISA observation bias} \label{app:bias}

%%%%%%%%%%%%%%%%%%%%%%%%%%%%%%%%%%%
\begin{figure*}
        \centering
	 \includegraphics[width=0.48\textwidth]{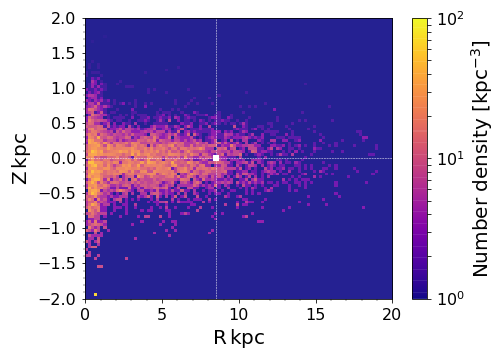} 
	 \includegraphics[width=0.48\textwidth]{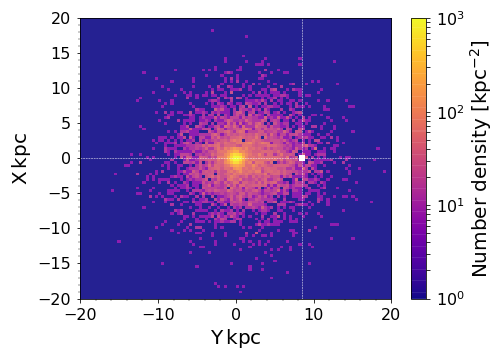}
	 \includegraphics[width=0.48\textwidth]{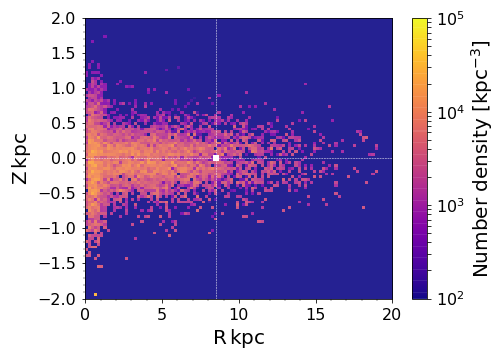} 
	 \includegraphics[width=0.48\textwidth]{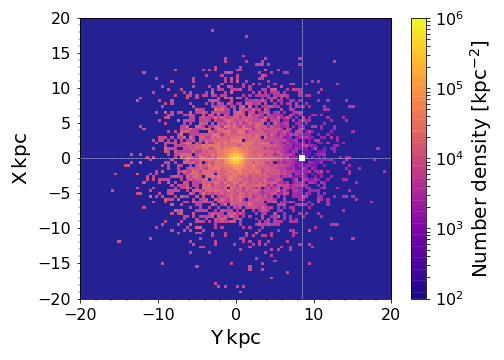}
         \caption{Top panels: Number density distribution of the DWDs detected by LISA in the Galactic $R-Z$ plane and in the Galactic equatorial plane $X-Y$. Bottom panels: same number density distributions corrected for the observation bias as described in Sect. \ref{app:bias}. A white square marks the position of the Sun.}
       \label{fig:7}
\end{figure*}
%%%%%%%%%%%%%%%%%%%%%%%%%%%%%%%%%%%%%

To derive a simplified analytic expression for the LISA observational bias, we assume that all DWD binaries have roughly the same chirp mass.
Indeed, the observed distribution of chirp masses is expected to range between 0.2 and 1$\,$M$_{\odot}$  \citep[see][fig.12]{Korol2017}.
Under this assumption, the SNR is only a function of distance $d$ and frequency $f$, and can thus be written as \citep[e.g.,][]{Maggiore}:
\begin{equation}
{\rm SNR} = K   \frac{f^{2/3}\sqrt{T_{\rm obs}/S_{\rm n}(f)}}{d} \equiv \frac{{\cal R}}{d},
\end{equation}
where $K$ is a constant that depends on the detector geometry, the sky location of the source, its orientation and chirp mass, and $S_{\rm n}$ is the noise spectral density of the detector.
At low frequencies ($10^{-4} - 10^{-2}\,$Hz) the noise spectral density scales as $S_{\rm n} \propto 1/f^{\alpha}$ where $\alpha \simeq 4.7$, as obtained by  fitting the LISA noise curve from \citet{LISA2017}. 
Therefore, ${\cal R} \sim f^{2/3 + \alpha/2}$ and 
\begin{equation}
\frac{d N}{d{\cal R}} =\frac {dN}{df}\frac{df}{d{\cal R}} = {\cal R}^{-(20+3\alpha)/(4+3\alpha)},
\end{equation}
where we have used the fact that the number of sources $N$ per frequency interval scales as $dN/df \propto f^{-11/3}$. This follows from
assuming that the population is in a steady state, i.e. that DWDs have a uniform distribution in time to merger \cite[e.g.][]{colacino}. 
By definition, a binary will be detected if observed with ${\rm SNR} = {\cal R}/d > 7$, so we can compute the LISA detection fraction as\footnote{Note that this expression is valid only at large distances $d$, because the steady-state distribution $dN/df \propto f^{-11/3}$ and the approximation $S_{\rm n} \propto 1/f^{\alpha}$ only hold
in a limited range of frequencies. This is also obvious from the fact that the detection fraction, $w$, diverges as $d\to 0$. In practice, however, eq.~(\ref{eqw}) reproduces well the results of our simulations.}
\begin{equation}\label{eqw}
w \propto \int^{+\infty}_{7d} \frac{dN}{d{\cal R}}d{\cal R} \approx F \, d^{-0.9}\,,
\end{equation}
with $F=\,$const.

We test this analytic expression using our mock  population.
We selected binaries with ${\rm SNR} > 7$ and bin them in the $R-\theta$ space, and we compare this to the same histogram without the cut in SNR.
The ratio between the two histograms represents the LISA detection fraction.
Next, we average the detection fractions over $\theta$ to express them as a function of $R$ only. 
We then fit the obtained detection fractions with  $w = F\, d^{\beta}$, and obtain $F = 0.016 \pm 0.04$ and  $\beta = 0.93 \pm 0.04$, consistent with the value in eq.~(\ref{eqw}).
To show the effect of the correction we compute the surface number density maps of DWDs in the Galactic $X-Y$ and $Z_R$ planes. In the top panels of Fig. \ref{fig:7} we show DWD density maps not corrected for the bias. In the bottom panels we show the same maps corrected for the bias by assigning a weight $w$ (evaluated using $F=0.016$ and $\beta = 0.93$) to each bin. The effect of the correction is clearly visible in the bottom right panel of Fig. \ref{fig:7}, where there are fewer sources around the Sun with respect to the upper right panel.

%%%%%%%%%%%%%%%%%%%%%%%%%%%%%%%%%%%%%%%%%%%%%%%%%%

\bsp	
\label{lastpage}
\end{document}